\documentclass[11pt,a4paper,english,twoside]{article}
\usepackage{a4wide}
\usepackage{amssymb, amsmath, amsthm}
\usepackage{graphicx}
\usepackage{subcaption}
\usepackage[all]{xy}
\usepackage[pdftex,hyperref,svgnames]{xcolor}
\usepackage[pdftex,colorlinks=true,
pdfstartview=FitV,
pdfnewwindow=true,
linktoc = page,
linkcolor= blue,
citecolor= red,
urlcolor= blue,
hyperindex=true,
hyperfigures=false]{hyperref}
\hypersetup{linktocpage}
\usepackage{dsfont}
\usepackage{empheq}
\usepackage{cite}
\usepackage{float}
\usepackage{cancel}
\usepackage{relsize}
\usepackage{soul}
\usepackage{enumerate}
\usepackage{enumitem}
\usepackage{hhline}
\usepackage{multirow}
\usepackage{xspace}
\usepackage{bbm}
\usepackage{pdfpages}
\usepackage{setspace}

\newcommand{\nn}{\nonumber}
\newcommand{\w}{\wedge}

\newcommand{\f}[2]{f^{#1}{}_{#2}}

\newcommand{\ib}{{\sl sugra} basis\xspace}
\newcommand{\pb}{{\sl geometry} basis\xspace}

\usepackage[normalem]{ulem}

\def\del {\partial}
\def\d {{\rm d}}
\def\mmm {\mathcal{M}}

\begin{document}
\numberwithin{equation}{section}

\begin{titlepage}
\begin{center}

\phantom{draft}

\vspace{2.0cm}

{\LARGE \bf{On classical de Sitter solutions\\[12pt] and parametric control}}\\[.5cm]

{David Andriot$^{1}$ and Fabian Ruehle$^{2,3}$}\\[1cm]
{\small\slshape $^1$ Laboratoire d'Annecy-le-Vieux de Physique Th\'eorique (LAPTh),\\
CNRS, Universit\'e Savoie Mont Blanc (USMB), UMR 5108,\\
9 Chemin de Bellevue, 74940 Annecy, France}\\[.2cm]
{\small\slshape $^2$ Department of Physics and Department of Mathematics,\\
Northeastern University, Boston, MA 02115, USA}\\[.2cm]
{\small\slshape $^3$ NSF Institute for Artificial Intelligence and Fundamental Interactions,\\
Boston, USA}\\[.5cm]
{\upshape\ttfamily andriot@lapth.cnrs.fr; f.ruehle@northeastern.edu}\\[1.8cm]
\end{center}

\begin{abstract}
Finding string backgrounds with de Sitter spacetime, where all approximations and corrections are controlled, is an open problem. We revisit the search for de Sitter solutions in the classical regime for specific type IIB supergravity compactifications on group manifolds, an under-explored corner of the landscape that offers an interesting testing ground for swampland conjectures. While the supergravity de Sitter solutions we obtain numerically are ambiguous in terms of their classicality, we find an analytic scaling that makes four out of six compactification radii, as well as the overall volume, arbitrarily large. This potentially provides parametric control over corrections. If we could show that these solutions, or others to be found, are fully classical, they would constitute a counterexample to conjectures stating that asymptotic de Sitter solutions do not exist. We discuss this point in great detail.
\end{abstract}

\end{titlepage}

\tableofcontents

\newpage

\section{Introduction and summary of results}\label{sec:intro}

Superstring theories in a ten-dimensional (10d) spacetime are known to admit, in a classical regime, effective theories which are the 10d supergravities. The classical regime refers to a double expansion in the string coupling constant $g_s$, and in the string length $l_s$ captured by the parameter $\alpha'= l_s^2$. The former requires $g_s < 1$ to provide a perturbative expansion in string loops; the latter requires $l_s/r < 1$ where $r$ corresponds to the typical length probed, or equivalently $1/r$ corresponds to the typical energy below the cut-off scale of the effective theory. The condition $l_s/r < 1$ allows both to truncate the string spectrum to its massless states and to restrict the effective theory to two-derivative terms, as the first level in a corresponding expansion. The smaller $g_s$ and $l_s/r$ are, the better the approximation using 10d supergravities is: corrections become negligible or at least controlled. In that case, a solution to 10d supergravity is a classical string background.

A crucial point, however, is that 10d supergravities are theories in their own right. Hence, although a supergravity theory would have a Planck mass, and $p$-brane charges, their relation to $l_s$ is a string theory input. Similarly, the relation of $g_s$ to the vev of the dilaton field in a given solution comes from string theory; at the supergravity level, a constant dilaton can actually be absorbed in a redefinition of the fluxes and $p$-brane charges, and disappears from the equations. Therefore, given a supergravity solution, asking whether it is a classical string background requires to make the $l_s$- and $g_s$-dependence explicit by relating it to an effective field theory derived from string theory. This is achieved on the one hand by relating the 10d Planck mass to $\alpha'$ and $g_s$, but also by imposing the ``source quantization'', i.e.~explicitly deriving the brane and orientifold charges as well as their amount from a string theory background, and similarly realizing flux quantization in units of $\alpha'$ (those two requirements can sometimes be combined into the so-called tadpole condition). In other words, the question of ``classicality'' of a supergravity solution, usually phrased as having a small $g_s$ and a large volume (referring to the typical size $r$ of compact extra dimensions, as compared to $l_s$), also requires flux and source quantization. In our setup, we need to add another requirement, which is an additional quantization condition dubbed the lattice condition, which ensures compactness of the curved extra dimensions. These five requirements were listed and studied for classical de Sitter solutions in \cite{Andriot:2020vlg}.

Determining the precise value of $g_s$ or $l_s/r$ for a typical radius $r$ can only be done on a case-by-case basis for a given supergravity solution. Instead, one often looks for a scaling parameter $\gamma> 1$, such that the solution can be scaled to another solution via $r\, \rightarrow\, \gamma\, r$ and $g_s\, \rightarrow \, \gamma^{-q}\, g_s$, $q>0$. If this is possible for an arbitrarily large $\gamma$, then $g_s$ and $l_s/r$ can be as small as desired, leading to parametric control (on classicality); the precise values of $g_s$ and $l_s/r$ in a given solution then do not matter. Note that the other requirements for classicality, e.g.~flux quantization, should remain compatible with such a scaling, for instance making an integer grow and not diminish. Such a situation has been realized in the seminal example of the so-called DGKT solution \cite{DeWolfe:2005uu} (see also \cite{Lust:2004ig, Camara:2005dc, Acharya:2006ne}), for which the 10d spacetime is a 4d anti-de Sitter spacetime, times a 6d torus orbifold. Thanks to flux quantization, one can show that the typical squared radius for the torus orbifold goes as $\kappa^{\frac{1}{3}} \, v_i \sim l_s^2 \, n^{\frac{1}{2}}$ and $g_s \sim n^{-\frac{3}{4}}$, with a (possibly large) parameter $n$ (see e.g.~\cite{Andriot:2023fss} for a precise evaluation).

Since $g_s$ and $r$ may also get mapped to scalar fields in a further 4d effective description, the existence of such a scaling is sometimes also investigated through a corresponding scalar potential. For de Sitter solutions, which are the subject of this paper, this has been the case in \cite{Roupec:2018mbn, Junghans:2018gdb, Banlaki:2018ayh, Grimm:2019ixq, Cicoli:2021fsd}. In those works and others \cite{Andriot:2019wrs, Andriot:2020vlg}, it was concluded that classical de Sitter solutions with parametric control likely do not exist (although loopholes were also indicated in these arguments). If this is true, the best one could hope for is a classical de Sitter solution with ``local'' or ``numerical'' control. The latter would mean that quantities such as $g_s$ and $l_s/r$, which control the corrections to the solution, are numerically small, without having necessarily a scaling parameter to further adjust them. In absence of any parameters this amounts to finding dimension zero loci (i.e., isolated points) in the parameter space. One could also have a ``local'' scaling parameter, meaning that the parameter is only allowed to vary in a finite range (and cannot be sent to infinity) \cite{Andriot:2020vlg}. The solution could then be viewed as located on an isolated patch in parameter space, not connected to the asymptotics. In those cases, the de Sitter solution would happen to be classical, meaning corrections are controlled but not parametrically. Note that in either case one has to carefully evaluate the quantities entering the classicality constraints, which requires technicalities associated to the five requirements listed above (flux quantization, etc.).

This line of thought for de Sitter solutions is in agreement with the latest versions of the swampland de Sitter conjecture \cite{Bedroya:2019snp, Rudelius:2021oaz, Rudelius:2021azq}, which claim that a de Sitter solution cannot be obtained in a 4d string effective theory (as a critical point in a positive scalar potential) at the asymptotics of field space. Interpreting the latter as the limit of small $g_s$ and $l_s/r$, one deduces indeed the absence of parametrically controlled de Sitter solutions in a classical regime. Nevertheless, a classical de Sitter solution in the bulk of field space, numerically or locally controlled, are a priori still allowed by these conjectures.

What is actually the situation when looking at concrete solutions? So far, most candidates for classical string background solutions with a 4d de Sitter spacetime have been obtained from type II supergravity compactifications on 6d group manifolds. A motivation for using those manifolds is that they are easy to handle (since they are essentially torus fibrations), while providing negative 6d curvature, ${\cal R}_6 <0$, a common requirement for such solutions in a smeared limit. A recent classification of such solutions, together with an extensive search for supergravity solutions, as well as a study of their properties (including stability), has been conducted in \cite{Andriot:2022way, Andriot:2022yyj, Andriot:2022bnb}. We refer to those, as well as to \cite{Andriot:2019wrs}, for a review and a list of known solutions. As explained previously, having de Sitter supergravity solutions is only a first step towards a classical string background. To test the latter, one needs to fulfill five requirements. The geometry of group manifolds complicates the concrete implementation of such tests. The authors of \cite{Roupec:2018mbn, Junghans:2018gdb, Banlaki:2018ayh} studied concrete examples in type IIA supergravity with $O_6$ orientifold planes (class $s_{6666}^+$ of \cite{Andriot:2022way}) and concluded that no parametric control with an appropriate scaling exists there. In \cite{Andriot:2020vlg} (see also \cite{Andriot:2020wpp}), a detailed evaluation of the relevant quantities for the five requirements was performed for certain type IIB supergravity solutions with $O_5$ planes (class $s_{55}^+$); we follow to some extent that reference here. The conclusion was also negative: while flux, source and lattice quantization conditions could be satisfied together with $g_s < 1$, it turned out that 2 of the 6 radii were substringy, namely $r_3, r_6 < l_s$. In a conservative view, this does not allow classicality and control over corrections, even though we will revisit such a conclusion in this paper. To summarize, no clear classical de Sitter string background has been identified up to now, not even one with local or numerical control rather than parametric control over corrections.

\medskip

In this paper, we revisit this problem in several ways. In section \ref{sec:solsetupgal}, we start by working-out in great details the formulation of the problem. We search for supergravity de Sitter solution in type IIB supergravity with $O_5$ planes in the class $s_{55}^+$. These obey a simple ansatz, for which some solutions are already known, in particular solution 14 \cite{Andriot:2020wpp}, denoted $s_{55}^+ 14$, obtained on the 6d group manifold with algebra $\mathfrak{g}_{3.5}^{0} \oplus \mathfrak{g}_{3.5}^{0}$. This solution is first found and expressed in terms of a set of variables \texttt{var1} (flux components, etc.), as in \eqref{sol14var1}. In order to identify the lattice conditions and ensure compactness, but also to quantize harmonic components of fluxes, we need to re-express it in terms of a second set of variables \texttt{var2}, going through a change of basis in the 6d manifold. Finally, we introduce a third set of variables \texttt{var3}, that is subject to the classicality constraints: those are the 6 radii $r_{1,...,6}, g_s$, flux integers, number of sources, etc. While the supergravity solutions are most easily found in terms of \texttt{var1}, the classicality constraints are easily expressed in terms of \texttt{var3}, which complicates the problem. The definitions and relations between these variables are given in section \ref{sec:solsetupgal}: an overview is given in section \ref{sec:firstglimpse} while a summary can be found in section \ref{sec:sum}. Technicalities on the change of basis and the quantization conditions leading to the classicality constraints are detailed in section \ref{sec:2basis} and \ref{sec:quant}, and appendix \ref{ap:sources}.

This detailed formulation then allowed us to launch advanced numerical searches for classical solutions. Three methods are presented in section \ref{sec:strategies}. The first one in section \ref{sec:2steps} is similar (though more systematic in its implementation) to the one used in \cite{Andriot:2020vlg}, and allows to provide values of \texttt{var3} for solution 14, in particular the radii $r_{1,...,6}$ and $g_s$ \eqref{sol14var3}. We note a missed factor of $2\pi$ in that reference, that we correct here, providing new values. An (a priori) more efficient method is proposed in section \ref{sec:directsearch}: there, we attempt direct searches of solutions in terms of \texttt{var3}, allowing to impose in the search the classicality constraints. However, it seems that we face there numerical difficulties, for reasons we explain and illustrate. A last method is proposed in section \ref{sec:constrsugrasearch}, leading to a new de Sitter solution $s_{55}^+ 29$, given in \eqref{sol29var1} and \eqref{sol29var3}. Qualitatively, it is similar to $s_{55}^+ 14$. While having the lattice, source and flux quantization conditions satisfied, the two solutions have
\begin{align}
\text{Solution 14:} \quad & g_s \approx 0.64\ ,\ r_{1,2,4,5} > l_s \ ,\  r_3 \approx 0.05\, l_s \ ,\ r_6 \approx 0.09 \, l_s\ ,\\
\text{Solution 29:} \quad & g_s \approx 0.53\ ,\ r_{1,2,4,5} > l_s \ ,\  r_3 \approx 0.07\, l_s \ ,\ r_6 \approx 0.08 \, l_s\ .
\end{align}
While $g_s<1$ and $r_{1,2,4,5} > l_s$ is satisfied, having $r_{3,6}< l_s$ is worrisome for classicality. It is actually non-trivial to conclude from this that the solution is  non-classical, and we discuss this point and related corrections in the Outlook; in particular, we note that $r_3, r_6$ normalize non-closed one-forms, and in that sense, do not correspond to volumes of one-cycles. Still, our conservative criteria for classicality are not all satisfied. In addition, we cannot find better solutions in that respect, and it is unclear to us whether there is a theoretical obstruction or simply a numerical limitation of finding the solutions.

Interestingly for the question of classicality, we present in section \ref{sec:scaling} a scaling with parameter $\gamma$, that maps one solution to another: the effect on the supergravity variables amounts to a homogeneous scaling with $1/\gamma^2$ in the 10d equations, ensuring that the equations remain solved. This $\gamma$-scaling, expressed on the variables \texttt{var3} \eqref{scaling}, has several appealing features: one is that it goes {\sl partially} towards the direction of a classical regime. In fact, there is a family of scalings with some freedom in how the individual quantities are scaled. In a simple setup that fixes some of the exponents, we get
\begin{equation}
r_{4,5} \rightarrow  \gamma \, r_{4,5} \ ,\quad  r_{1,2} \rightarrow \gamma^{\frac{1}{2}}\, r_{1,2} \ , \quad g_s, r_{3}, r_6 \ \text{invariant}\,.
\end{equation}
Four of the radii then grow parametrically while the other two as well as $g_s$ are invariant. This results in parametric control over the volume of the compactification manifold. We also note that the scaling is such that the classicality constraints are preserved, up to a discretization of $\gamma$ due to its action on a flux integer (which grows with $\gamma$ in this case). Therefore, if we had a solution with admissible values for $g_s, r_{3}, r_{6}$, we could claim for the first time a classical de Sitter solution with parametric control over classicality, analogously to the DGKT solution mentioned above.

We present this scaling at the level of the 10d theory in section \ref{sec:scaling10d}. In section \ref{sec:4dscaling} and appendix \ref{ap:changebasis}, we attempt to identify it in a corresponding 4d theory, using the work of \cite{Andriot:2022bnb}. The 4d scalar potential, as well as the 4d Ricci scalar ${\cal R}_4$ in 4d Einstein frame, are expected to scale as $1/\gamma^5$ (at least in simple versions of the $\gamma$-scaling). We face difficulties to realize the $\gamma$ scaling at the 4d level via a combined action on the 10d background quantities entering the 4d scalar potential and the 4d scalar fields. One reason is the lack of an explicit map of 10d quantities such as the radii to the 4d fields, especially when including off-diagonal 6d metric fluctuations. Despite this situation in 4d, we still discuss the question of scale separation in section \ref{sec:parcontrol} and conclude that it cannot be achieved using the scaling in our de Sitter solutions. Indeed, we identify a Kaluza--Klein tower that scales in the same way as ${\cal R}_4$, which means that the mass gap is not parametrically large. A more detailed analysis of this question including other contributions to the spectrum is beyond the scope of this work. In 10d on the other hand, a classical de Sitter solution with parametric control would be possible if we found solutions with appropriate values for $g_s, r_{3}, r_{6}$. Perhaps even the solutions we have now do not receive stringy contributions, due to the fact that the substringy objects are not cycles as mentioned above and corrections need to be evaluated.

The scaling freedom seems to contradict the swampland claims that asymptotic de Sitter solutions do not exist. The lack of clear numerical control on $r_{3}, r_{6}$ and on the corrections, as well as the difficulties in getting a 4d realization of the scaling, means that we do not have a conclusive counter-example to these claims. Nevertheless, in section \ref{sec:parcontrol}, we revisit the arguments against such solutions put forward in \cite{Junghans:2018gdb,Grimm:2019ixq, Cicoli:2021fsd}. We identify the precise loopholes in those arguments and find that our scaling and setup actually utilizes these loopholes. A major reason is that our compactification space is a solvmanifold with curvature \textsl{and} a specific fibration structure. More precisely, the geometry is described by pairs of structure constants. Compactness only quantizes their product, not their values individually. This allows to balance the magnitude of one versus the other, in conjunction with adjusting the radii; we refer to section \ref{sec:parcontrol} for a more precise discussion. This property of solvable algebras and solvmanifolds is a building block in the $\gamma$ scaling, and helps in getting these asymptotic solutions.

\medskip

At this stage, one may wonder whether a classical de Sitter solution of the type described above could be used for cosmology. Focusing on the universe today and the question of dark energy, we believe that such solutions are interesting candidates that would deserve more investigation. Solutions 14 and 29 are perturbatively unstable, with $\eta_V \sim -4$ due to a single tachyonic field direction;\footnote{Such a perturbative instability makes these de Sitter solutions agree with the refined de Sitter swampland conjecture \cite{Garg:2018reu, Ooguri:2018wrx}. The latter is an older version of the conjecture, when compared to the latest claim, mentioned above, of no asymptotic de Sitter solution, on which we focus in this work.} dark energy would thus not be realized via the common scenario of a cosmological constant due to a positive potential minimum. But it turns out that a quintessence scenario, with a field rolling down from a de Sitter maximum, even with $\eta_V \sim - {\cal O}(1)$, can actually be in agreement with observations, as pointed out e.g.~in \cite{Agrawal:2018rcg}. Indeed, the key point is that the Hubble friction could hold a field on top of such a maximum, until a recent time where the Hubble parameter got lowered to match the value of that potential maximum, allowing the field to start rolling. We also note that models considering a varying equation of state parameter $w = w_0 + w_a (1-a)$ get the observational constraints $w_0 \sim -0.77, w_a \sim -0.83$ \cite{DES:2024tys}. The fact that $w_0 >-1$ and $w_a<0$ signals the possibility of starting at a de Sitter point where $w=-1$ in a recent past, and gain some kinetic energy until today. While requiring a more detailed analysis, such possibilities with the solutions above are appealing. We also note that the scaling freedom naturally allows for solutions with a small cosmological constant. More ideas for future investigations are presented in an outlook in section \ref{sec:outlook}.

\section{Classical de Sitter solutions: setup and variables}\label{sec:solsetupgal}

In this section, we present the formalism needed to search for classical de Sitter solutions. The search itself is described in section~\ref{sec:strategies}. Beyond the general compactification setup, we present the variables in terms of which equations should be expressed to find solutions. Three different sets of variables will be introduced, going from variables of a supergravity solution (e.g.~a flux component) to those with which classicality of the solution can be tested (e.g.~a flux integer, the string coupling and radii). An overview is first provided in subsection \ref{sec:firstglimpse} while a summary is given in subsection \ref{sec:sum}. Technicalities are detailed in subsections \ref{sec:2basis} and \ref{sec:quant}, together with appendix \ref{ap:sources}.

\subsection{Supergravity solution variables and quantized versions: a first glimpse}\label{sec:firstglimpse}

We look for de Sitter solutions of 10d type IIB supergravity that can obey the requirements for being a classical string background. Starting with supergravity, we consider solutions with a 10d spacetime being the direct product of a 4d de Sitter spacetime and a 6d compact manifold $\mmm$, where the latter is a group manifold. In the 6d (internal) dimensions, we work with 1-forms $e^a = e^{a}{}_m \d y^m$, $a=1,...,6$ called coframes. As $\mmm$ is a group manifold, the underlying Lie algebra structure constants $f^a{}_{bc}$ are related to the spin connection on $\mmm$. The 1-forms obey the Maurer-Cartan equation
\begin{equation}
\d e^a = - \frac{1}{2}\, f^a{}_{bc}\, e^b \w e^c \ . \label{MCeq}
\end{equation}
The $f^a{}_{bc}$ encode most of the 6d geometry, together with the 6d metric $g_{ab}=\delta_{ab}$, which is trivial in this basis. Such de Sitter solutions on group manifolds have been classified in \cite{Andriot:2022way}, and we focus here on the class $s_{55}^+$. These solutions include sources that are orientifold $O_5$-planes and $D_5$-branes. We arrange them into three sets (labeled by $I=1,2,3$) of parallel sources as follows:
\begin{align}
\begin{tabular}{|l|c|c|c|c|c|c|}
        \hline
         & \multicolumn{6}{|c|}{Internal Dimension $a$} \\
        \hhline{~------}
        Source Set $I$ &&&&&& \\[-11pt]
        & 1 & 2 & 3 & 4 & 5 & 6 \\
        \hline
        \hline
        $I=1$ ($D_5$~\text{and}~$O_5$) & $\times$ & $\times$ & & & & \\
        \hline
        $I=2$ ($D_5$~\text{and}~$O_5$) & & & $\times$ & $\times$ & & \\
        \hline
        $I=3$ ($D_5$~\text{only}) & & & & & $\times$ & $\times$ \\
        \hline
\end{tabular}
\end{align}
We denote the contributions of each source set $I$, i.e.,~their charges and amount, by $T_{10}^I$.

The orientifold projections put restrictions on the allowed structure constants and flux components. The latter are denoted $H_{abc}$ for the NSNS $H$-flux 3-form in the $e^a$ basis, and $F_{1\,a}, F_{abc}$ for the components of the $F_1$ and $F_3$ RR fluxes; we will not consider $F_5$ fluxes. The field content (structure constants, flux components, source contributions) allowed by the orientifold projections in the class $s_{55}^+$ is listed in \cite[(2.14)]{Andriot:2022way} and we will consider part of it in the following by restricting the class to a specific ansatz. We refer to \cite{Andriot:2022way} for more detail on the general compactification and solution ansatz. In particular, the dilaton $\phi$ is constant and captured by $g_s= e^{\phi}$. There is also no warp factor and the $O_5/D_5$ sources are smeared. A consequence of this ansatz is that all variables appearing in 10d equations, namely
\begin{equation}
\delta_{ab},\ f^a{}_{bc},\ H_{abc},\ g_s F_{q\, a_1..a_q},\ g_s T_{10}^I \ , \label{var}
\end{equation}
are constant, which helps solving the 10d supergravity equations. In addition to \cite{Andriot:2022way}, the following is also based on \cite{Andriot:2020wpp, Andriot:2020vlg, Andriot:2022yyj}.

One difficulty with group manifolds, especially solvmanifolds, is to understand their global aspects, in particular compactness. A compact group manifold $\mmm$ is constructed by taking the quotient of a Lie group $G$ by a discrete subgroup $\Gamma\subset G$, $\mmm=G/\Gamma$. The action of $\Gamma$ leads to global identifications of coordinates of the manifold associated to $G$, rendering $\mmm$ compact. The existence of such a lattice is not always known or guaranteed for solvable groups. Moreover, when such a lattice exists, its explicit action is often not easy to implement. The situation we can control is when the 1-forms $e^a$ that solve Maurer-Cartan equation~\eqref{MCeq} can be expressed explicitly in terms of coordinates. In this case, the global identifications of the coordinates (the lattice action) leave the $e^a$ invariant up to quantization conditions of the structure constants, as we will illustrate in examples in \eqref{eag3.5bis} and \eqref{eag3.4}. Such a controlled situation is typically realized when we are in a basis $\{ {e^a}' \}$ where only a few structure constants are non-zero, which allows us to solve~\eqref{MCeq}. However, supergravity solutions are found more conveniently in a basis $\{ e^a \}$ with many non-zero structure constants. We refer to the bases $\{ e^a \}$ and $\{ {e^a}' \}$ as the \ib and \pb, respectively. The degrees of freedom get shuffled around in the two bases: in the \ib, we have a trivial metric $g=\mathbbm{1}$, and typically many non-zero $f^a{}_{bc}$. The simple form of the metric helps to solve the supergravity equations. We can then change to the \pb via an isomorphism described by a matrix $M$, $e' =M e$, such that only few of the $f^a{}_{bc}'$ are non-zero at the cost of having a more complicated metric $g = M^{-T} \mathbbm{1} M^{-1}$. This metric now contains the degrees of freedom previously encoded in the structure constants. In the \pb, we can work out the quantization conditions of the structure constants, and understand more generally the global aspects of the 6d geometry. Both bases are therefore necessary, and we will construct $M$ to the variables~\eqref{var} in both bases. We refer to the variables in the \ib as \texttt{var1}.

So far we have only discussed the supergravity solution. In order to test its ``classicality'', we need to quantize fluxes and source terms. Importantly, the quantization of fluxes only needs to be imposed on their harmonic representative. Irrespective of the basis, any $p$-form can be Hodge-decomposed into a harmonic part, an exact part, and a co-closed part. For example for $F_3$,
\begin{equation}
F_3 = F_{3\, {\rm harmo}} + \d A + *_6 \d B \ , \label{Hodgedecompo}
\end{equation}
where $F_{3\, {\rm harmo}}, A, B$ are globally defined forms. Together with the rest of the variables in the \pb such as the structure constants, metric, and source contributions, this data forms the set \texttt{var2}. This set of variables will be the one on which all quantization conditions can be imposed and classicality can be tested.

To give an example of how this works, let us consider a part of the 3-form flux given by $F_{\omega}\;\omega$, where $F_{\omega}$ is constant and $\omega$ is a harmonic form for a cycle $\sigma$ with radius $r$. The flux quantization condition then requires
\begin{equation}
\int_{\sigma }\omega = (2\pi r)^3 \ , \quad \frac{1}{(2\pi l_s)^2} \int_\sigma F = N_{\omega} \in \mathbb{Z} \quad \Rightarrow \quad F_{\omega} = \frac{1}{2\pi l_s} \, N_{\omega} \left(\frac{l_s}{r}\right)^3 \ .
\end{equation}
Note that the radius dependence enters since in our conventions, the 1-forms $e^a$ contain the radii but the metric does not. Consequently, the flux components contain inverse radii (they can be viewed as multiplied by vielbeins). On top of integers encoding the flux quantization conditions, we get further integers from source term quantization (which also involves some volumes and radii), and from the structure constants quantization conditions. All these integers, the radii, $g_s$, and a few more quantities enter in the last set of variables necessary to describe a solution, which we call \texttt{var3}. In the following, we will detail relations between the various variable sets and how they enter the classicality constraints.

\subsection{Supergravity solution ansatz in the two bases and harmonic flux components}\label{sec:2basis}

In this subsection, we define different sets of variables, in particular \texttt{var1} and \texttt{var2}, and give explicit relations between them. Those are needed in the solution search and the classicality study. This part is somewhat technical and readers can refer to the summary in subsection~\ref{sec:sum}.

We want to search for de Sitter solutions in terms of the constant variables~\eqref{var}. We further restrict the ansatz by setting some of these variables to zero, and keep only the following ones:
\begin{align}
\texttt{var1}:\quad & f^{1}{}_{45}, f^{1}{}_{46}, f^{2}{}_{35}, f^{2}{}_{45}, f^{2}{}_{46}, f^{3}{}_{15}, f^{3}{}_{25}, f^{6}{}_{14} \label{var1}\\
& F_{1\, 5} \nn\\
& F_{135}, F_{136}, F_{146}, F_{235}, F_{236}, F_{246} \nn\\
& H_{125}, H_{346} \nn \\
& T_{10}^1, T_{10}^2, T_{10}^3\,, \nn
\end{align}
where in particular $F_5=0$ and we dropped the subscript $3$ for the components of $F_3$. This set of variables, \texttt{var1}, contains quantities in the \ib. A further restriction to our ansatz is the requirement that four of the structure constants are non-zero
\begin{equation}
f^{1}{}_{46}\, f^{6}{}_{14}\, f^{2}{}_{35}\, f^{3}{}_{25} \neq 0 \ . \label{fabcneq0}
\end{equation}
As we will see, this is the ansatz of de Sitter solutions 14 and 15 \cite{Andriot:2020wpp, Andriot:2020vlg} and of solutions 22-27 of \cite{Andriot:2021rdy}; a complete solution database can be found in \cite{Andriot:2022way} or \cite{Andriot:2022bnb}.

Some properties of these solutions can be anticipated already by solving some of 10d equations (listed in \cite[App.~B.2]{Andriot:2022way}) in terms of the fields~\eqref{var1}. To start with, the only non-trivial Jacobi identity that the structure constants need to satisfy fixes $f^{2}{}_{46}$ in terms of other structure constants
\begin{equation}
f^{2}{}_{46} = - \frac{f^{3}{}_{15}}{f^{3}{}_{25}} f^{1}{}_{46} \ . \label{relfabc}
\end{equation}
Turning to fluxes, the $H$-flux Bianchi identity $\d H=0$ leads to the relation
\begin{equation}
H_{125} = - \frac{f^{3}{}_{25}}{f^{1}{}_{46}} H_{346} \ .\label{relH}
\end{equation}
Introducing the parameter $b= \frac{H_{125}}{f^{3}{}_{25}}$, one can show that\footnote{ We use the notation  $e^{125} = e^1 \w e^2 \w e^5$, etc.}
\begin{equation}
H= b \, (f^{3}{}_{25}\, e^{125} - f^{1}{}_{46}\, e^{346} ) = \d \left( b \left(e^{13} + \frac{f^{1}{}_{45}}{f^{2}{}_{35}} e^{24}  \right)  \right) \ ,
\end{equation}
where we used the Maurer-Cartan equation \eqref{MCeq}. This means that $H$ has to be an exact 3-form. Turning to $F_3$ and using our assumption $F_5=0$, its equation of motion becomes
\begin{equation}
\d *_6 F_3 = 0 \ .\label{F3eom}
\end{equation}
This means that the components of $F_3$ satisfy
\begin{equation}
\f{2}{35} F_{135} - \f{1}{46} g^{12} F_{146} - \f{1}{46} F_{246} = 0 \ .\label{relF3}
\end{equation}

As a consequence (using \eqref{fabcneq0}), $F_3$ has only 5 independent components. As for $H$, which is exact, we would like to perform a Hodge decomposition~\eqref{Hodgedecompo} for $F_3$. We can use the fact that $F_3$ is even under an $O_5$ projection. This restricts $A$ and $B$,
\begin{align}
F_3 & = \sum_i  F_{3\, \omega_i}\, \omega_i \\
& +  \d  \left( c_{12}\, e^{12} + c_{34}\, e^{34} + c_{56}\, e^{56} \right) + *_6 \d  \left( a_{12}\, e^{12} + a_{34}\, e^{34} + a_{56}\, e^{56} \right) \ , \label{F3decompo}
\end{align}
where we sum over harmonic forms $\omega_i $, to be specified. It is straightforward to verify that \eqref{F3eom}, together with \eqref{fabcneq0}, implies the absence of an exact part in $F_3$, namely $c_{12}= c_{34} = c_{56}=0$. We will be more explicit about the other components of $F_3$ in the following. But we can indicate already that the five independent components of $F_3$ correspond to the five independent parameters $ F_{3\, \omega_1},  F_{3\, \omega_2},  a_{12} ,  a_{34} , a_{56}$.

\medskip

The solution ansatz \eqref{var1}-\eqref{fabcneq0} for the variables \texttt{var1} allows for an analytic change of basis to the \pb $\{ {e^a}' \}$, where we have better control over the 6d geometry. The change of basis $e' = M e$ is expressed in terms of three real parameters that we denote (for reasons that will become clear) by $g^{12}, g^{34}, g^{56}$,
\begin{equation}
{e^{a\neq {2,3,6}}}' = e^a \ , \ {e^2}'= e^2 + g^{12}\, e^1 \ ,\ {e^3}'= e^3 + g^{34}\, e^4 \ ,\ {e^6}'= e^6 + g^{56}\, e^5 \ . \label{changebasis} \\
\end{equation}
As mentioned previously, this leads to a change of the metric from the identity $\delta_{ab}$ in the \ib to $g_{ab}$ in the \pb. One has $g^{-1} = M \mathbbm{1}^{-1} M^T$, i.e.
\begin{equation}
g^{-1}= \, \left( \begin{array}{cccccc} 1 & g^{12} & & & & \\ g^{12} & 1+ (g^{12})^2 & & & & \\  & & 1 + (g^{34})^2 & g^{34} & & \\  & & g^{34} & 1 & & \\  & & & & 1 & g^{56} \\  & & & & g^{56} & 1+ (g^{56})^2  \end{array} \right) \ .\label{g-1}
\end{equation}
This explains the notation of the three real parameters: they are off-diagonal components of the inverse metric. In terms of those, one also gets the metric
\begin{equation}
g= \, \left( \begin{array}{cccccc} 1+ (g^{12})^2 & -g^{12} & & & & \\ -g^{12} & 1 & & & & \\  & & 1 & -g^{34} & & \\  & & -g^{34} & 1 + (g^{34})^2 & & \\  & & & & 1+ (g^{56})^2  & -g^{56} \\  & & & & -g^{56} & 1 \end{array} \right) \ .\label{gab}
\end{equation}
This change of basis and metric was found and discussed for solution 14 in \cite[App. C]{Andriot:2020wpp}.

More concretely, we fix the three parameters as follows in terms of the initial structure constants,
\begin{equation}
g^{12}= \frac{\f{3}{15}}{\f{3}{25}} \, , \ g^{34} = \frac{\f{1}{46}\f{2}{45}- \f{1}{45}\f{2}{46}}{\f{1}{46}\f{2}{35}} \, , \ g^{56}=\frac{\f{1}{45}}{\f{1}{46}} \ .\label{gijinv}
\end{equation}
Doing so allows to set four structure constants in the \pb to zero, as can be verified with the Maurer-Cartan equation~\eqref{MCeq}. The remaining four structure constants are not modified by this change of basis, so we do not need label them with a prime: those are the non-zero ones in \eqref{fabcneq0},
\begin{equation}
f^{2}{}_{35}, f^{3}{}_{25}, f^{1}{}_{46}, f^{6}{}_{14} \ .
\end{equation}
This change of basis then achieves what was explained above: it reduces the structure constants to a few that are non-zero, from which we can get an explicit and global description of the 6d geometry. As indicated in \eqref{relfabc}, $f^{2}{}_{46}$ can be fixed in terms of other structure constants. Therefore, three independent structure constants are set to zero by the change of basis. The information they carry goes into the three parameters $g^{12}, g^{34}, g^{56}$, which appear in the \pb as metric components. The inverse relation, from \eqref{gijinv} and \eqref{relfabc}, reads
\begin{equation}
\f{3}{15} = g^{12} \f{3}{25} ,\ \f{1}{45} = g^{56} \f{1}{46} ,\ \f{2}{46} = - g^{12} \f{1}{46} ,\ \f{2}{45} = g^{34} \f{2}{35} - g^{12} g^{56} \f{1}{46}  \ .
\end{equation}

Let us look at the transformation of the various quantities under this change of basis. An important feature is that this change of basis preserves the volumes of the three source sets (see related discussions in \cite[App. A]{Andriot:2022way} and \cite[App. C]{Andriot:2020wpp}), namely ${e^{12}}' = e^{12}$, ${e^{34}}' = e^{34}$, ${e^{56}}' = e^{56}$. As a consequence, the source contributions
\begin{equation}
T_{10}^1, \ T_{10}^2, \ T_{10}^3 \ ,
\end{equation}
are the same in both bases and we do not need to worry about the orientifold projections. Turning to fluxes, we first note that $F_1$ is unchanged: indeed, it has a single component along the $a=5$ direction, and $e^5={e^5}'$, so its component is not modified
\begin{equation}
F_1= F_{1\, 5}\, e^5 = F_{1\, 5}\, {e^5}' \ .
\end{equation}
Moreover $e^5$ is a harmonic form in either basis.

The first flux that changes is the $H$-flux. Its expression as an exact form is the same in both bases,
\begin{equation}
H = \d \left( b \left(e^{13} + \frac{f^{1}{}_{45}}{f^{2}{}_{35}} e^{24}  \right)  \right) =  \d \left( b \left({e^{13}}' + \frac{f^{1}{}_{45}}{f^{2}{}_{35}} {e^{24}}'  \right)  \right) \ ,
\end{equation}
because the 2-forms that appear above get modified by a closed form. The $H$-flux receives a new contribution in the \pb, for which we get the relations
\begin{equation}
H_{125}' = H_{125} = b f^3{}_{25} \ ,\ H_{346}' = H_{346} = - b f^1{}_{46}\ ,\ H_{345}'= b f^1{}_{46} g^{56} \ .
\end{equation}

Finally, the change in the $F_3$ components is more involved. The components in the \pb (including new ones) are given as follows
\begin{align}
& F_{135}' =    F_{135} - g^{56} F_{136} - g^{12} F_{235} +     g^{12} g^{56} F_{236}, \nn \\
&  F_{145}' = F_{145} -g^{34} F_{135} + g^{34} g^{56} F_{136}  - g^{56} F_{146} + g^{12} g^{34} F_{235} -  g^{12} g^{34} g^{56} F_{236}\nn\\
& \qquad \quad \ - g^{12} F_{245} + g^{12} g^{56} F_{246}, \nn\\
&  F_{136}' = F_{136} - g^{12} F_{236}, \nn\\
&  F_{146}' = F_{146} -g^{34} F_{136} +  g^{12} g^{34} F_{236} - g^{12} F_{246}, \label{Fprime} \\
&  F_{235}' = F_{235} - g^{56} F_{236}, \nn\\
&  F_{245}' = F_{245} -g^{34} F_{235} + g^{34} g^{56} F_{236}  - g^{56} F_{246},\nn\\
&  F_{236}' = F_{236}, \nn\\
&  F_{246}' = F_{246} -g^{34} F_{236}  \ .\nn
\end{align}

To summarize, in order to get a better control on the 6d geometry, we perform a change from the \ib to the \pb, given by \eqref{changebasis} and \eqref{gijinv} on the one-forms. The resulting changes are schematically
\begin{align}
& e^a \rightarrow {e^a}' \nn\\
& \delta_{ab} \rightarrow g_{ab} \nn\\
& H_{abc} \rightarrow H_{abc}' \nn\\
& F_{abc} \rightarrow F_{abc}' \ , \label{changebasissummary}
\end{align}
while $F_{1\, 5} = F_{1\, 5}'$ is unchanged, the $T_{10}^I$ are unchanged, and out of the structure constants, four remain non-zero and are unchanged. The new metric is not diagonal anymore, and some flux components are changed.

\medskip

We now come back to the expressions of fluxes in terms of harmonic forms, in order to obtain the set of variables \texttt{var2}. Recall that the $H$-flux is exact, so it does not need quantization, while $F_1$ is already expressed as a harmonic form. Thus, we focus on $F_3$ decomposed as in \eqref{F3decompo}. We identified in \cite{Andriot:2020vlg} four harmonic 3-forms with constant components in the co-frame basis, $\omega_{i=1,...,4}$. We recall that $\omega_3 = {e^{156}}' + \d o_3$, $\omega_4 = {e^{234}}' + \d o_4$, where $o_3$ and $o_4$ are proportional to ${e^{36}}'$. These two harmonic forms are odd under the $O_5$ projections, so they do not contribute to $F_3$. We thus focus on $\omega_{1,2}$, and rewrite them in both bases as follows
\begin{align}
& \omega_1 = {e^{146}}' + \d o_1 = e^{146} + \d \tilde{o}_1 \\
& \omega_2 = {e^{235}}' + \d o_2 = e^{235} + g^{12} e^{135} + \d \tilde{o}_2 \nn\\
{\rm with}\ \ & \tilde{o}_1 = \alpha \f{1}{46} \Big(  e^{12} + \frac{\f{1}{46}}{\f{3}{25}} g^{56} e^{34} \Big) - \frac{\alpha \f{1}{46} \f{2}{35}}{\f{6}{14}} g^{34} e^{56} \nn\\
 & \tilde{o}_2 = \alpha \f{2}{35} \Big( - e^{12}  + \frac{\f{2}{35}}{\f{6}{14}} g^{34} e^{56} \Big) - \frac{\alpha \f{2}{35}\f{1}{46}}{\f{3}{25}} g^{56}  e^{34} \nn\\
 & \alpha= \frac{ g^{12}  }{(\f{2}{35})^2 + (\f{1}{46})^2 (1 + (g^{12})^2)} \ .\nn
\end{align}
In the \ib, this leads to
\begin{align}
\omega_1 & = e^{146} + \alpha \f{1}{46} \Big(\f{2}{35}  e^{135} - \f{1}{46} ( e^{246} + g^{12} e^{146}) \Big) \\
& = \alpha \f{1}{46} \f{2}{35}  e^{135}  + \frac{\alpha}{g^{12}} \Big((\f{2}{35})^2 + (\f{1}{46})^2 \Big) e^{146} - \alpha  (\f{1}{46})^2 e^{246} \nn\\
\omega_2 & =  e^{235} + \alpha (\f{1}{46})^2 (1 + (g^{12})^2) e^{135}  + \alpha \f{1}{46} \f{2}{35}  ( e^{246} + g^{12} e^{146}) \nn
\end{align}
It is easy to verify that they are harmonic in the \ib, since $\d \omega_i = \d *_6 \omega_i = 0$. We furthermore find that
\begin{align}
*_6 \d  \Big( a_{12}\, e^{12} & + a_{34}\, e^{34} + a_{56}\, e^{56} \Big) = - a_{12} \f{1}{46} e^{135} +  (a_{12} \f{1}{45} - a_{34} \f{3}{25} ) e^{136} \nn\\
& - a_{12} \f{2}{46} e^{235} + (a_{12} \f{2}{45} + a_{34} \f{3}{15} + a_{56} \f{6}{14} ) e^{236} - a_{12} \f{2}{35} e^{246}   \ .
\end{align}
Expanding $F_3$ as
\begin{equation}
F_3 = \sum_{i=1,2}  F_{3\, \omega_i}\, \omega_i  + *_6 \d  \left( a_{12}\, e^{12} + a_{34}\, e^{34} + a_{56}\, e^{56} \right) \ ,
\end{equation}
which is the same in either basis, we can now relate its components to those in the \ib
\begin{align}
& F_{135} = F_{3\, \omega_1} \alpha  \f{1}{46} \f{2}{35} + F_{3\, \omega_2} \alpha  (\f{1}{46})^2 (1 + (g^{12})^2) - a_{12} \f{1}{46} \\
& F_{136} = a_{12} \f{1}{45} - a_{34} \f{3}{25} \nn\\
& F_{146} = F_{3\, \omega_1} \frac{\alpha}{g^{12}} \Big((\f{2}{35})^2 + (\f{1}{46})^2 \Big)  + F_{3\, \omega_2} \alpha \f{1}{46} \f{2}{35} g^{12} \nn\\
& F_{235} = F_{3\, \omega_2} + a_{12} \f{1}{46} g^{12} \nn\\
& F_{236} = a_{12} \f{2}{45} + a_{34} \f{3}{15} + a_{56} \f{6}{14}  \nn\\
& F_{246} = -F_{3\, \omega_1} \alpha (\f{1}{46})^2 + F_{3\, \omega_2} \alpha \f{1}{46} \f{2}{35} - a_{12} \f{2}{35} \nn \ .
\end{align}
A cross-check for these expressions is that \eqref{relF3} is satisfied. It is also useful to invert the above, to get
\begin{align}
& F_{3\, \omega_1} = \left(1+ \alpha g^{12} (\f{1}{46})^2 (1 + (g^{12})^2) \right)\, F_{146} - \alpha \f{1}{46} \f{2}{35} g^{12}\, (F_{235} + g^{12} F_{135}) \ ,\\
& F_{3\, \omega_2} = \frac{\alpha}{g^{12}} \Big((\f{2}{35})^2 + (\f{1}{46})^2 \Big)\, (F_{235} + g^{12} F_{135}) - \alpha \f{1}{46} \f{2}{35} g^{12}\, F_{146} \nn  \ .
\end{align}
We finally introduce the set of variables \texttt{var2}
\begin{align}
\texttt{var2}:\quad & f^{1}{}_{46}, f^{2}{}_{35}, f^{3}{}_{25}, f^{6}{}_{14}, g^{12}, g^{34}, g^{56} \label{var2}\\
& F_{1\, 5} \nn\\
& F_{3\omega_1}, F_{3\omega_2}, a_{12}, a_{34}, a_{56} \nn\\
& b \nn \\
& T_{10}^1, T_{10}^2, T_{10}^3 \nn
\end{align}
These are the quantities in the \pb, expressed in terms of harmonic components for the fluxes. This is the variable set on which we will be able to apply all quantization conditions, in the next subsection. We also gave explicit relations between \texttt{var1} and \texttt{var2}.

\subsection{Quantization conditions, 6d algebras and classicality}\label{sec:quant}

We are now ready to give the quantization conditions for the variables~\eqref{var2} in the set \texttt{var2}, namely the flux quantization of their harmonic components, the source quantization in terms of their number, and the lattice conditions on the structure constants. The latter depend on the 6d algebra underlying the group manifold and we will detail the possible algebras given our solution ansatz. Finally, given these quantization conditions, we will list the constraints to get a classical string background. This will involve a third set of variables which we call \texttt{var3}. Again, readers may skip this technical information and read the summary in subsection~\ref{sec:sum}.

The advantage of the \pb is that the group manifold geometry can be understood more easily. One can find explicit expressions for the one-forms ${e^a}'$ in terms of coordinates, and read off the fibration structure and the global coordinate identifications related to the lattice. We will do this following mostly~\cite{Andriot:2020wpp,Andriot:2020vlg}. A first application is to get an expression for the structure constants in the \pb in the form
\begin{equation}
{f^a{}_{bc}}' = \frac{r_a\, N_{a}}{r_b r_c} \ , \label{fabcNa}
\end{equation}
where the numbers $N_a$ are quantized according to lattice conditions and $r_a > 0$ are ``radii'' appearing in the expressions of ${e^a}'$. We recall that in our case, the non-zero structure constants in the \pb are equal to those in the \ib, so we drop the prime on them. Another benefit of the \pb is that it is easy to obtain the norm of harmonic forms, using their explicit expression as discussed in \cite{Andriot:2020vlg},
\begin{equation}
\int \omega_1 = \int {e^{146}}' = (2 \pi)^3\, r_1 r_4 r_6 \ ,\ \int \omega_2 = \int {e^{235}}' = (2 \pi)^3\, r_2 r_3 r_5 \ .
\end{equation}
This norm is useful to impose string theory flux quantization, which requires for a harmonic U(1) flux on a cycle $\Sigma$
\begin{equation}
\frac{1}{(2\pi l_s)^{q-1}} \int_{\Sigma} F_q = N_q \in \mathbb{Z} \ ,
\end{equation}
where $l_s$ is the string length. Finally, the string origin of the source contributions fixes their charge and number as
\begin{equation}
\frac{T_{10}^I}{p+1} = (2^{p-5} N_{O_p}^I - N_{D_p}^I) \,  \frac{(2\pi l_s)^{7-p}}{\textit{vol}_{\bot_I}} \ , \label{T10I}
\end{equation}
with integers $N_{O_p}^I$ and $N_{D_p}^I$ being the number of $O_p$ and $D_p$ in the set $I$. The transverse volume $\textit{vol}_{\bot_I} \equiv \int {\rm vol}_{\bot_I}$ could in principle be computed in the \pb. However, as explained in more detail in Appendix~\ref{ap:sources}, this can be subtle since in the 10d spacetime, spaces wrapped are not always along cycles, even though the volume forms are always globally defined. We take $\textit{vol}_{\bot_I} = (2 \pi)^{9-p}\ r_{a_{1\,\bot_I}} \ldots r_{a_{9-p\,\bot_I}}$, corresponding to schematically counting radii as $2\pi r_a \equiv \int {e^a}'$.

Finally, as discussed in \cite[Sec. 4.2]{Andriot:2020wpp}, there exists an overall scaling symmetry of all 10d equations,
\begin{equation}
T_{10}^I \rightarrow \frac{1}{\lambda^2}\, T_{10}^I \ ,\quad F_q,\, H,\, f^a{}_{bc} \rightarrow \frac{1}{\lambda}\, F_q,\, H,\, f^a{}_{bc} \ ,\ \quad \lambda > 0 \ . \label{lambdascaling}
\end{equation}
Thus, we can rescale any given supergravity solution prior to imposing the quantization conditions. This freedom can certainly help, so we introduce the parameter $\lambda$ as an extra variable in the previous relations. It will play a further crucial role in section~\ref{sec:scaling}.

\medskip

The above, applied to our solution ansatz, leads us to consider the following expressions or quantization conditions for the relevant variables in \texttt{var2} in \eqref{var2}, to ensure a string origin and a compact group manifold
\begin{align}
& \frac{1}{\lambda}\, g_s F_{1\, 5}=\frac{1}{2\pi l_s} \times \frac{g_s\, N_{1\, 5}\, l_s}{r_5} \ ,\ \frac{1}{\lambda}\, g_s F_{3\, \omega_1}= \frac{1}{2\pi l_s} \times \frac{g_s\, N_{\omega_1} \, l_s^3}{r_1 r_4 r_6}  \ ,\ \frac{1}{\lambda}\, g_s F_{3\, \omega_2}= \frac{1}{2\pi l_s} \times \frac{g_s\, N_{\omega_2}\, l_s^3}{r_2 r_3 r_5} \ , \nn\\
& \frac{1}{\lambda^2}\, g_s T_{10}^1 =\frac{1}{(2\pi l_s)^2} \times \frac{6\, g_s\, N_{s1} \, l_s^4}{r_3 r_4 r_5 r_6}\ ,\  \frac{1}{\lambda^2}\, g_s T_{10}^2 =\frac{1}{(2\pi l_s)^2} \times \frac{6\, g_s\, N_{s2} \, l_s^4}{r_1 r_2 r_5 r_6}\ , \nn\\  &\frac{1}{\lambda^2}\, g_s T_{10}^3 =\frac{1}{(2\pi l_s)^2} \times \frac{6\, g_s\, N_{s3} \, l_s^4}{r_1 r_2 r_3 r_4}\ , \nn\\
& \frac{1}{\lambda}\, f^2{}_{35} =\frac{1}{2\pi l_s} \times \frac{r_2\, N_{2} \, 2\pi l_s}{r_3 r_5} \ ,\ \frac{1}{\lambda}\, f^3{}_{25} =\frac{1}{2\pi l_s} \times \frac{r_3\, N_{3} \, 2\pi l_s}{r_2 r_5} \ , \nn\\
& \frac{1}{\lambda}\, f^1{}_{64} =\frac{1}{2\pi l_s} \times \frac{r_1\, N_{1} \, 2\pi l_s}{r_4 r_6} \ , \ \frac{1}{\lambda}\, f^6{}_{14} =\frac{1}{2\pi l_s} \times \frac{r_6\, N_{6} \, 2\pi  l_s}{r_1 r_4} \ , \label{quantizedqtties0}
\end{align}
where we introduced the source numbers $N_{sI}=N_{O_5}^I - N_{D_5}^I$. In the following, we drop the $2\pi l_s$ factors, as they can be absorbed in a redefinition of $\lambda$. This amounts to giving the supergravity quantities in units of $2\pi l_s$. In addition, each $r_a/l_s$ will be replaced by $r_a$, by which we mean that radii are in given in units of $l_s$. This leads to the following simplified relations
\begin{align}
& g_s F_{1\, 5}= \frac{g_s\, \lambda\, N_{1\, 5}}{r_5} \ ,\ g_s F_{3\, \omega_1}=  \frac{g_s\, \lambda\, N_{\omega_1}}{r_1 r_4 r_6}  \ ,\ g_s F_{3\, \omega_2}= \frac{g_s\, \lambda\, N_{\omega_2}}{r_2 r_3 r_5} \ ,  \nn\\
& g_s T_{10}^1 = \frac{6\, g_s\, \lambda^2\, N_{s1}}{r_3 r_4 r_5 r_6}\ ,\  g_s T_{10}^2 =\frac{6\, g_s\, \lambda^2\, N_{s2}}{r_1 r_2 r_5 r_6}\ , \  g_s T_{10}^3 = \frac{6\, g_s\, \lambda^2\, N_{s3}}{r_1 r_2 r_3 r_4}\ , \label{quantizedqtties} \\
& f^2{}_{35} = \frac{r_2\, \lambda\, N_{2}\, 2\pi}{r_3 r_5} \ ,\ f^3{}_{25} =\frac{r_3\, \lambda\, N_{3} \, 2\pi}{r_2 r_5} \ , \ f^1{}_{64} =\frac{r_1\, \lambda\, N_{1} \, 2\pi}{r_4 r_6} \ , \ f^6{}_{14} =  \frac{r_6\, \lambda\, N_{6} \, 2\pi}{r_1 r_4} \ .\nn
\end{align}
Let us emphasize that the $ 2\pi$ in the structure constants {\it has been forgotten} in \cite{Andriot:2020vlg}, which will lead to slightly different results in the following. We also used different sign conventions for the structure constants.

The relations \eqref{quantizedqtties} lead us to introduce a last set of variables
\begin{align}
\texttt{var3}:\quad  & N_1,\ N_2,\ N_3,\ N_6,\ N_{1\, 5},\ N_{\omega_1},\ N_{\omega_2},\ N_{s1},\ N_{s2},\ N_{s3}, \nn\\
& \ r_{a=1,...,6},\ g_s,\ \lambda ,\nn\\
& g^{12},\ g^{34},\ g^{56},\ a_{12},\ a_{34},\ a_{56}, \ b\ . \label{var3}
\end{align}
Note that $g_s$ had not been included in \texttt{var1} and \texttt{var2}. Numerical supergravity solutions are found in terms \texttt{var1}, where $F_q$ and $T_{10}^I$ are actually given by $g_s F_q$ and $g_s T_{10}^I$. In \texttt{var3}, we really extract a value for $g_s$, hence the need to list it explicitly.

These new variables come with constraints: We know already that the flux and source numbers $N_{1\, 5}, N_{\omega_1}, N_{\omega_2}, N_{sI}$ are integers. In addition, the $N_{sI}$ are bounded from above by $N_{O_5}^I \leq 2^4= 16$ when the 4 transverse dimensions are circles (each with 2 fixed points), e.g.~in a torus and for most manifolds to be considered here. Finally, the structure constant numbers $N_a$ are bound to the lattice conditions. We summarize the constraints as
\begin{equation}
N_{1\, 5}, N_{\omega_1}, N_{\omega_2}, N_{s1}, N_{s2}, N_{s3} \in \mathbb{Z} \ ,\ N_{sI} \leq N_{O_5}^I \ ,\ N_a\,:\ \mbox{lattice conditions} \ .\label{constraintsquant}
\end{equation}

To specify the lattice conditions explicitly, we need a detailed knowledge of the 6d geometry. In the \pb, the four non-zero structure constants consist of two independent pairs
\begin{equation}
f^{2}{}_{35}\,,~f^{3}{}_{25},\qquad f^{1}{}_{46}\,,~f^{6}{}_{14} \ .
\end{equation}
It means that the underlying algebra is a direct sum of two 3-dimensional (solvable) algebras, and the same structure holds for the group manifold. Depending on signs, there are two possible 3-dimensional algebras (see e.g.~\cite{Andriot:2022yyj})
\begin{equation}
\mathfrak{g}_{3.4}^{-1}:~f^{2}{}_{35} f^{3}{}_{25} > 0  \,,\qquad \mathfrak{g}_{3.5}^{0}:~f^{2}{}_{35} f^{3}{}_{25} < 0  \,,
\end{equation}
and the same holds for $f^{1}{}_{64}\,f^{6}{}_{14}$. Referring to solutions studied in \cite{Andriot:2020wpp,Andriot:2020vlg}, solution 14 is on $\mathfrak{g}_{3.5}^{0} \oplus \mathfrak{g}_{3.5}^{0}$ while solution 15 is on $\mathfrak{g}_{3.4}^{-1} \oplus \mathfrak{g}_{3.4}^{-1}$. Solutions 22-27 of \cite{Andriot:2021rdy} were found on $\mathfrak{g}_{3.4}^{-1} \oplus \mathfrak{g}_{3.5}^{0}$, with $\mathfrak{g}_{3.4}^{-1}$ corresponding to the pair $f^{2}{}_{35},\, f^{3}{}_{25}$. Note that a fourth combination is a priori possible, since the two pairs are not equivalent in the compactification, given the placement of sources. However, as we will explain below, we will not consider this option. The choice of algebras determines the lattice conditions, and the group manifold, ensuring its compactness as mentioned in subsection \ref{sec:firstglimpse}.

All 1-forms in the \pb obey the Maurer-Cartan equation \eqref{MCeq}. Prior to the $\lambda$ rescaling and the use of stringy units, we gave the expression of structure constants in \eqref{fabcNa}. This gives the equations
\begin{align}
& \!\!\!\!\! \d {e^2}' = -\frac{N_2 r_2}{r_3 r_5} \ {e^{35}}' \ , \ \d {e^3}' = - \frac{N_3 r_3}{r_2 r_5} \  {e^{25}}' \ ,\ \d {e^5}'=0 \ , \\
& \!\!\!\!\! \d {e^1}' = -\frac{N_1 r_1}{r_4 r_6}\ {e^{64}}' \ , \ \d {e^6}' = - \frac{N_6 r_6}{r_1 r_4}\ {e^{14}}' \ ,\ \d {e^4}'=0 \ , \nn
\end{align}
where we recall that we use different conventions with respect to \cite{Andriot:2020vlg}. Thanks to the simplicity of these equations in the \pb, we can solve them explicitly in terms of coordinates. Depending on the algebra, we get different solutions, with different global completions.

We first consider the pair with directions 1,4,6, defining the algebra $\mathfrak{g}_{3.5}^{0}$. One then has $f^{1}{}_{64} f^{6}{}_{14} < 0$, implying $N_1 N_6 <0$. Considering for instance $f^{6}{}_{14} < 0$, i.e.~$N_6 <0$, one solution to the Maurer-Cartan is given by\footnote{Similar one-forms are used in \cite[(8)-(10)]{Grana:2013ila}, which, however, absorbs the radii $r_1, r_4, r_6$ in the coordinates (the latter are said to be real and possibly rescaled). Later in \cite[Sec. 2.2]{Grana:2013ila}, the authors restrict their analysis to unit radii prior to conducting the lattice analysis. In our one-form parametrization, $r_1, r_4, r_6$ appear explicitly and the coordinates are given by the angular part only, which is the appropriate generalization. It leads to the same matrix $A$ and same quantization results~\eqref{lattices}.}
\begin{align}
{e^6}' & = r_6  \left( \cos(\sqrt{|N_1 N_6|} y^4)\, \d y^6 - \left|\frac{N_6}{N_1}\right|^{\frac{1}{2}} \sin(\sqrt{|N_1 N_6|} y^4)\, \d y^1  \right) \nn\\
{e^1}' & = r_1  \left( \left|\frac{N_1}{N_6}\right|^{\frac{1}{2}} \sin(\sqrt{|N_1 N_6|} y^4)\, \d y^6 + \cos(\sqrt{|N_1 N_6|} y^4)\, \d y^1  \right) \nn\\
{e^4}' & = r_4\, \d y^4 \ .\label{eag3.5bis}
\end{align}
The range of coordinates is $y^m \in [0, 2\pi]$, with periodic identification. The normalization is such that ${e^{16}}'= r_1r_6 \, \d y^1 \w \d y^6$, which after integration agrees with the normalization we chose for the quantization conditions. Further arguments from Appendix \ref{ap:sources} justify choosing that solution. We now define the corresponding (squashed) rotation matrix
\begin{equation}
A(y^4) = \left(\begin{array}{cc} \cos(\sqrt{|N_1 N_6|} y^4) &  - \left|\frac{N_6}{N_1}\right|^{\frac{1}{2}} \sin(\sqrt{|N_1 N_6|} y^4) \\ \left|\frac{N_1}{N_6}\right|^{\frac{1}{2}} \sin(\sqrt{|N_1 N_6|} y^4) & \cos(\sqrt{|N_1 N_6|} y^4) \end{array} \right)  \ .\label{rotbis}
\end{equation}
As argued in \cite{Andriot:2010ju, Andriot:2020vlg}, the following extra identification is required to have globally defined 1-forms
\begin{equation}
\left(\!\! \begin{tabular}{c} $y^6$ \\ $y^1$ \end{tabular} \!\!\right)_{y^4 + 2 \pi} = A(-2 \pi)\, \left(\!\! \begin{tabular}{c} $y^6$ \\ $y^1$ \end{tabular} \!\!\right)_{y^4} \ ,
\end{equation}
together with possible extra shifts of the coordinates by multiples of $2\pi$. This is also the condition on the fibration in order for the manifold to be compact. This identification is admissible if $A(-2 \pi)$ is an integer matrix, leading to the three possibilities to consider
\begin{equation}
\sqrt{|N_1 N_6|} \in \mathbb{N}^*  \ ,\quad \sqrt{|N_1 N_6|} \in \mathbb{N} + \frac{1}{2} \ ,\quad \sqrt{|N_1 N_6|} \in \mathbb{N} + \frac{1}{4} \ {\rm and}\ N_6=-N_1 \ .\label{lattices}
\end{equation}
These are the possible lattice conditions, which give rise to different global geometries. The first option corresponds to having topologically a torus $T^3$ with, a priori, a non-Ricci flat metric. The two other are globally more involved geometries, still with a torus cover. We refer to \cite{Andriot:2020vlg} for more details and references. The last two lattice options ease the search for classical solutions, but complicate the computation of volumes and of the number of fixed points, i.e., the number of orientifolds. So we will stick with the first (torus) lattice condition $\sqrt{|N_1 N_6|} \in \mathbb{N}^*$. The number of orientifold in each set is then that of a transverse torus, which is 16 in this case. In practice, we will take
\begin{equation}
N_6=-\frac{1}{N_1} \ .\label{latticecond}
\end{equation}

We turn to the second pair with directions 2,3,5, that we take to define the algebra $\mathfrak{g}_{3.4}^{-1}$, with $f^{2}{}_{35} f^{3}{}_{25} > 0$, i.e.~$N_2 N_3 >0$. The individual signs of $N_i$ do not need to be specified. One solution to the Maurer-Cartan equation is
\begin{align}
{e^2}' & = r_2 \left( \cosh(\sqrt{N_2 N_3} y^5)\, \d y^2  + \left(\frac{N_2}{N_3}\right)^{\frac{1}{2}} \sinh(\sqrt{N_2 N_3} y^5)\, \d y^3  \right) \nn\\
{e^3}' & = r_3 \left( \left(\frac{N_3}{N_2}\right)^{\frac{1}{2}} \sinh(\sqrt{N_2 N_3} y^5)\, \d y^2 + \cosh(\sqrt{N_2 N_3} y^5)\, \d y^3  \right) \nn\\
{e^5}' & = r_5\,  \d y^5 \ .\label{eag3.4}
\end{align}
The normalization gives ${e^{23}}' = r_2 r_3 \, \d y^2 \w \d y^3$, again in agreement with previous normalizations. We should now consider the (weighted) ``hyperbolic rotation'' matrix
\begin{equation}
\tilde{A}(y^5) = \left(\begin{array}{cc} \! \cosh(\sqrt{N_2 N_3} y^5) & \!\!\!\! \left(\frac{N_2}{N_3}\right)^{\frac{1}{2}} \sinh(\sqrt{N_2 N_3} y^5) \\ \! \left(\frac{N_3}{N_2}\right)^{\frac{1}{2}} \sinh(\sqrt{N_2 N_3} y^5) & \!\!\!\! \cosh(\sqrt{N_2 N_3} y^5) \end{array} \right) \nn
\end{equation}
and require as before, for globally defined 1-forms, to have the identification (together with possible shifts)
\begin{equation}
\left(\!\! \begin{tabular}{c} $y^2$ \\ $y^3$ \end{tabular} \!\!\right)_{y^5 + 2 \pi} = \tilde{A}(-2 \pi)\, \left(\!\! \begin{tabular}{c} $y^2$ \\ $y^3$ \end{tabular} \!\!\right)_{y^5} \ ,
\end{equation}
where the matrix has to take integer values. This leads to the following lattice conditions \cite{Andriot:2010ju}
\begin{align}
& \cosh(\sqrt{N_2 N_3} 2\pi) = n_1\ ,\ n_1^2 - n_2n_3=1\ , \ \frac{N_3}{N_2}=\frac{n_3}{n_2} \ ,\quad n_{1,2,3} \in \mathbb{N}^* \\
\leftrightarrow\quad  & N_2= \pm \frac{{\rm arccosh} (n_1)}{2\pi} \sqrt{\frac{n_2}{n_3}} \ , \ N_3= \pm \frac{{\rm arccosh} (n_1)}{2\pi} \sqrt{\frac{n_3}{n_2}} \ ,\ n_1^2 - n_2n_3=1  \ ,\quad n_{1,2,3} \in \mathbb{N}^* \nn
\end{align}
Note that one simple choice that solves the above is $n_3=1$, which guarantees that $n_2= n_1^2 - 1$ is integer. We can then express quantities in terms of $n_1>1$ only,
\begin{equation}
N_2= \pm \frac{{\rm arccosh} (n_1)}{2\pi} \sqrt{n_1^2-1} \ , \ N_3= \pm \frac{{\rm arccosh} (n_1)}{2\pi} \frac{1}{\sqrt{n_1^2-1}} \ ,\ n_{1} \in \mathbb{N}^*\ ,\ n_1>1 \ .\label{N2N3simple}
\end{equation}
Another solution for $n_1>1$ is $n_2=n_1-1, n_3=n_1+1$, from which we can again express $N_2, N_3$. Despite these simplifications, we will never manage to find a realization of these complicated ``hyperbolic'' lattice conditions in the supergravity solutions considered; we already failed in doing so in \cite{Andriot:2020vlg} for solution 15. One reason could be that for given real numbers $N_2,N_3$, it is unlikely to find an integer $n_1$ that would satisfy the simple relations \eqref{N2N3simple}. For example, when solving $\cosh(2\pi\sqrt{N_2 N_3}) = n_1$ numerically on a computer, the argument of $\cosh$ becomes a rational number due to finite precision. However, the cosh of a rational (non-zero) number is transcendental by the Lindemann-Weierstrass theorem and hence cannot equal an integer $n_1$. We will give more details on the numerical difficulty of this problem in section \ref{sec:strategies}. As a consequence, we will further restrict our ansatz to
\begin{equation}
f^{2}{}_{35}\,f^{3}{}_{25} <0,\qquad f^{1}{}_{64}\,f^{6}{}_{14}<0 \ , \label{ff<0}
\end{equation}
which results in a 6d manifold corresponding to $\mathfrak{g}_{3.5}^{0} \oplus \mathfrak{g}_{3.5}^{0}$. This signals already the difficulty to test classicality of any other supergravity solution.

\medskip

Having determined the lattice conditions, which are essentially $N_1 N_6 =-1$, $N_2 N_3=-1$, and $N_{O_5}^I=16$, the constraints for the quantization conditions in \eqref{constraintsquant} are completely fixed. They ensure compactness of the 6d manifold, and are necessary requirements for a string origin of the solution. We are then left to specify criteria for the supergravity solution to be a classical string background, which we take to mean that
\begin{equation}
g_s \ll 1 \ ,\ r_a \gg 1 \,, \label{classicality}
\end{equation}
where the radii are given in string length units. For simplicity, we will impose ``$<$'' and ``$>$'' instead of ``$\ll$'' and ``$\gg$''. As will be discussed in the Outlook, the classicality conditions \eqref{classicality} might be too conservative choices for the radii, but we stick to this definition of a classical solution for now. Before looking for such solutions, let us summarize all the material presented so far.

\subsection{Summary: variables, relations and constraints}\label{sec:sum}

We have introduced the setup for supergravity solutions in section \ref{sec:firstglimpse}, and we have motivated the use of three different sets of variables and two different bases to check classicality of a solution. The technical details have been developed in subsections \ref{sec:2basis} and \ref{sec:quant}, which we summarize here. An overview is given in figure~\ref{fig:VarRelations}.

We start with an ansatz for a de Sitter solution in type IIB supergravity within the class~$s_{55}^+$. In our ansatz, we keep only the fields in \eqref{var1}, that form a set of 20 variables denoted \texttt{var1}. Known supergravity solutions are given in those variables. Our ansatz also assumes four non-zero structure constants, cf.~\eqref{fabcneq0}. This allows to show that three variables in \texttt{var1} are not independent of the others, as indicated in \eqref{relfabc}, \eqref{relH}, \eqref{relF3}. A first change of variables is then performed towards 17 other variables, denoted as the set \texttt{var2} in \eqref{var2}. Eight variables remain unchanged through this transformation. The change of variable serves two purposes: first, it goes from the \ib with variables \texttt{var1} to the \pb, where we get a better control over the 6d geometry, needed to ensure compactness through lattice conditions. Second, it identifies harmonic representatives for the flux components  (mostly $F_3$), which we need for the flux quantization conditions. We summarize this as follows:
\begin{align}
\texttt{var1}:\quad &\ T_{10}^1, \ T_{10}^2, \ T_{10}^3,\ f^{1}{}_{46},\ f^{2}{}_{35},\ f^{3}{}_{25},\ f^{6}{}_{14} ,\ F_{1\, 5},\\
& \xymatrix{f^{1}{}_{45},\ f^{2}{}_{45},\ f^{2}{}_{46},\ f^{3}{}_{15},\ F_{135}, F_{136}, F_{146}, F_{235}, F_{236}, F_{246}, H_{125}, H_{346} \ar[d]_{\eqref{newoldfields}} \\ T_{10}^1, \ T_{10}^2, \ T_{10}^3,\ f^{1}{}_{46},\ f^{2}{}_{35},\ f^{3}{}_{25},\ f^{6}{}_{14} ,\ F_{1\, 5}, \hspace{1.45in} \ar@<-10pt>[u]_{\eqref{oldnewfields}}}\nn\\
\texttt{var2}:\quad & \ g^{12},\ g^{34},\ g^{56},\ F_{3\omega_1},\ F_{3\omega_2},\ a_{12},\ a_{34},\ a_{56}, \ b \nn
\end{align}
where the change of variables is given by
\begin{align}
& \f{1}{45} = g^{56} \f{1}{46} ,\ \f{2}{45} = g^{34} \f{2}{35} - g^{12} g^{56} \f{1}{46} ,\ \f{2}{46} = - g^{12} \f{1}{46} ,\ \f{3}{15} = g^{12} \f{3}{25} \ ,\nn\\
& F_{135} = F_{3\, \omega_1} \alpha  \f{1}{46} \f{2}{35} + F_{3\, \omega_2} \alpha  (\f{1}{46})^2 (1 + (g^{12})^2) - a_{12} \f{1}{46} \nn\\
& F_{136} = a_{12}  \f{1}{46} g^{56} - a_{34} \f{3}{25} \nn\\
& F_{146} = F_{3\, \omega_1} \frac{\alpha}{g^{12}} \Big((\f{2}{35})^2 + (\f{1}{46})^2 \Big)  + F_{3\, \omega_2} \alpha \f{1}{46} \f{2}{35} g^{12} \nn\\
& F_{235} = F_{3\, \omega_2} + a_{12} \f{1}{46} g^{12}  \label{oldnewfields}\\
& F_{236} = a_{12} \Big( \f{2}{35} g^{34} -  \f{1}{46} g^{12} g^{56} \Big) + a_{34} \f{3}{25} g^{12}  + a_{56} \f{6}{14}  \nn\\
& F_{246} = -F_{3\, \omega_1} \alpha (\f{1}{46})^2 + F_{3\, \omega_2} \alpha \f{1}{46} \f{2}{35} - a_{12} \f{2}{35} \ ,\ {\rm with} \ \alpha= \frac{ g^{12}  }{(\f{2}{35})^2 + (\f{1}{46})^2 (1 + (g^{12})^2)} \ ,\nn\\
& H_{125} = f^{3}{}_{25} b \ ,\ H_{346} = - f^{1}{}_{46} b \ .\nn
\end{align}
and the inverse transformation is
\begin{align}
& g^{12}= \frac{\f{3}{15}}{\f{3}{25}} \, , \ g^{34} = \frac{\f{1}{46}\f{2}{45}- \f{1}{45}\f{2}{46}}{\f{1}{46}\f{2}{35}} \, , \ g^{56}=\frac{\f{1}{45}}{\f{1}{46}} \ ,\nn\\
& F_{3\, \omega_1} = \left(1+ \alpha g^{12} (\f{1}{46})^2 (1 + (g^{12})^2) \right)\, F_{146} - \alpha \f{1}{46} \f{2}{35} g^{12}\, (F_{235} + g^{12} F_{135}) \nn\\
& F_{3\, \omega_2} = \frac{\alpha}{g^{12}} \Big((\f{2}{35})^2 + (\f{1}{46})^2 \Big)\, (F_{235} + g^{12} F_{135}) - \alpha \f{1}{46} \f{2}{35} g^{12}\, F_{146} \nn\\
& a_{12}  = \frac{ F_{235} - F_{3\, \omega_2} }{\f{1}{46} g^{12}} = \dots \label{newoldfields}\\
& a_{34}  = a_{12}  \frac{\f{1}{46}}{\f{3}{25}} g^{56} - \frac{F_{136}}{\f{3}{25}} = \dots \nn\\
& a_{56}  =  \frac{F_{236}}{\f{6}{14}} - a_{12} \Big( \frac{\f{2}{35}}{\f{6}{14}} g^{34} -  \frac{\f{1}{46}}{\f{6}{14}} g^{12} g^{56} \Big) - a_{34} \frac{\f{3}{25}}{\f{6}{14}} g^{12} = \dots \nn\\
& b= \frac{H_{125}}{f^{3}{}_{25}} \ ,\nn
\end{align}
where the dots indicate that the resulting expressions can be evaluated further using the transformations given in the equations above; we refrain from doing this explicitly since the resulting equations are long and not very illuminating.

\begin{figure}[t]
\scriptsize\centering
\begin{tabular}{|p{.26\textwidth}|}
\hline
\begin{center}
\vspace{-.5cm}
\ib\\
\texttt{var1}
\vspace{-.3cm}
\end{center}\\
\hline
\hline
\vspace{-.3cm}
\begin{itemize}[leftmargin=*]
\item[$\checkmark$] Identity metric
\item[$\times$] Description of 6d geometry
\item[$\times$] Quantization conditions
\item[$\times$] Classicality check
\end{itemize}\vspace{-.5cm}\\
\hline
\end{tabular}
$\xleftrightarrow{\quad}$
\begin{tabular}{|p{.26\textwidth}|}
\hline
\begin{center}
\vspace{-.5cm}
\pb\\
\texttt{var2}
\vspace{-.3cm}
\end{center}\\
\hline
\hline
\vspace{-.3cm}
\begin{itemize}[leftmargin=*]
\item[$\times$] Identity metric
\item[$\checkmark$] Description of 6d geometry
\item[$\checkmark$] Quantization conditions
\item[$\times$] Classicality check
\end{itemize}\vspace{-.5cm}\\
\hline
\end{tabular}
$\xleftrightarrow{\quad}$
\begin{tabular}{|p{.26\textwidth}|}
\hline
\begin{center}
\vspace{-.5cm}
\pb\\
\texttt{var3}
\vspace{-.3cm}
\end{center}\\
\hline
\hline
\vspace{-.3cm}
\begin{itemize}[leftmargin=*]
\item[$\times$] Identity metric
\item[$\checkmark$] Description of 6d geometry
\item[$\checkmark$] Quantization conditions
\item[$\checkmark$] Classicality check
\end{itemize}\vspace{-.5cm}\\
\hline
\end{tabular}
\caption{Summary of the choice of bases, sets of variables, and their advantages ($\checkmark$) and disadvantages ($\times$).}
\label{fig:VarRelations}
\end{figure}

The set of variables \texttt{var2} is the correct one for imposing quantization conditions (flux quantization, source quantization, lattice quantization), which are necessary to ensure a stringy origin of the supergravity solution as well as compactness of the 6d manifold. Imposing these constraints leads to a new set of variables \texttt{var3} as in \eqref{var3}, subject to additional constraints. We also need to enforce the classicality conditions, which ensure that the supergravity solution is a classical string background. We summarize the above as follows
\begin{align}
\texttt{var2}:\quad &\ T_{10}^1, \ T_{10}^2, \ T_{10}^3,\ f^{1}{}_{46},\ f^{2}{}_{35},\ f^{3}{}_{25},\ f^{6}{}_{14} ,\ F_{1\, 5},\ F_{3\omega_1},\ F_{3\omega_2}, \nn\\
& \xymatrix{g^{12},\ g^{34},\ g^{56},\ a_{12},\ a_{34},\ a_{56}, \ b \hspace{1.6in}  \ar@{.>}[d]_{} \\ N_{s1},\ N_{s2},\ N_{s3},\ N_1,\ N_2,\ N_3,\ N_6,\ N_{1\, 5},\ N_{\omega_1},\ N_{\omega_2}, \hspace{0.3in} \ar@<-10pt>[u]_{\eqref{quantizedqtties}}}\nn\\
\texttt{var3}:\quad & \ r_{a=1,...,6},\ g_s,\ \lambda ,\nn\\
& g^{12},\ g^{34},\ g^{56},\ a_{12},\ a_{34},\ a_{56}, \ b\ .
\end{align}
The (simplified) relation between variables was given in \eqref{quantizedqtties}. The set \texttt{var3} contains more variables than \texttt{var2}, but the former are also subject to additional constraints,
\begin{align}
{\rm Constraints}:\quad & 0< g_s < 1 \ ,\ r_a > r_{min} = 1 \ ,\ \lambda >0  \label{constraints}\\
& N_{1\, 5}, N_{\omega_1}, N_{\omega_2} \in \mathbb{Z} \nn\\
& N_{s1}, N_{s2}, N_{s3} \in \mathbb{Z} \ ,\ N_{sI=1,2,3} \leq N_{O_5}^I =16 \nn\\
& N_6=-\frac{1}{N_1} \ ,\ N_3=-\frac{1}{N_2} \nn
\end{align}
where we fixed $N_{O_5}^I =16$, and we restricted ourselves to toroidal geometries as discussed around \eqref{latticecond} and \eqref{ff<0}. We also recall that $f^{2}{}_{35} f^{3}{}_{25} <0$ and $f^{1}{}_{64} f^{6}{}_{14}<0$. Other possibilities with more complicated lattice conditions have been discussed above. Finally, we fixed $r_{min} = 1$ but in practice, we will have to lower it, meaning that some radii will be below the string length. These extra constraints make it difficult to find an explicit set \texttt{var3} that satisfies all requirements.

We now have all the formalism necessary to search for supergravity solutions
that satisfy the constraints \eqref{constraints}, ensuring them to be
classical string backgrounds.

\section{Searches for classical solutions}\label{sec:strategies}

In this section, we present three methods or strategies followed to find classical de Sitter solutions, the difficulties encountered for each of them, and our results. As we will show, finding these solutions is a hard numerical optimization problem.

\subsection{The two-step procedure}\label{sec:2steps}

This first method is the most natural one, and has been used before. Here, we will adapt it and make it more systematic. The first step in this procedure is to solve all supergravity equations (equations of motion, Bianchi and Jacobi identities), listed in \cite[App. B]{Andriot:2022way}. This should be done with the constraint of having a de Sitter solution, namely ${\cal R}_4^S> 0$.\footnote{The constraint $T_{10}^I \leq 0$ for any set $I$ of sources containing no $O_p$ (but only $D_p$) needs to be imposed as well.} Prior to this search for a supergravity de Sitter solution, one should also specify the solution ansatz considered here. All this has been performed and automatized in previous works; the outcome is indeed supergravity de Sitter solutions, expressed in variables \texttt{var1} \eqref{var1}. More precisely, solution 14 on $\mathfrak{g}_{3.5}^{0} \oplus \mathfrak{g}_{3.5}^{0}$ and solution 15 on $\mathfrak{g}_{3.4}^{-1} \oplus \mathfrak{g}_{3.4}^{-1}$ were found this way in~\cite{Andriot:2020wpp} , and solutions 22-27 on $\mathfrak{g}_{3.4}^{-1} \oplus \mathfrak{g}_{3.5}^{0}$ in~\cite{Andriot:2021rdy}. The algorithm that implements the above strategy was improved and published as {\tt MSSS.nb} in~\cite{Andriot:2022way}, together with a solution database; there, solutions are referred to as $s_{55}^+ 14$, $15$, $22-27$. We refer to \cite{Andriot:2022way} for more details. For example, solution 14 in {\tt Solutions.nb} is
\begin{equation}\label{sol14var1}		
\begin{aligned}
&\text{Solution 14 (\texttt{var1}):}\quad {\cal R}_4^S = 0.022658 \ ,\\[8pt]
&g_s T_{10}^1= 10\ ,\ g_s T_{10}^2= -0.088507\ ,\ g_s T_{10}^3= -0.776520\ ,\\[6pt]
&g_s F_{1\,5}= 0.273982\ ,\ g_s F_{135}= 0.561224\ ,\ g_s F_{136}= -0.719988 \ ,\ g_s F_{146}= -0.052797 \ ,\\[6pt]
&g_s F_{235}= -0.677331\ ,\ g_s F_{236}= 0.313286\ ,\ g_s F_{246}= -0.178054\ ,\\[6pt]
& H_{125}= -0.004579 \ ,\ H_{346}= 0.228882\ ,\\[6pt]
&f^1{}_{45}= 0.843571 \ ,\ f^1{}_{46}= 0.671542 \ ,\ f^2{}_{35}= -0.289299 \ ,\ f^2{}_{45}= -0.061420\ ,\\[6pt]
&f^2{}_{46}= -0.810472\ ,\ f^3{}_{15}= 0.016213\ ,\ f^3{}_{25}= 0.013433 \ ,\ f^6{}_{14}= 0.413104\ .
\end{aligned}
\end{equation}

As a second step, referred to as the classicality check, one takes a concrete solution expressed in terms of \texttt{var1}, and performs the double change of variables \texttt{var1} $\rightarrow$ \texttt{var2} $\rightarrow$ \texttt{var3}. More precisely, one performs the change to \texttt{var2}, and then looks for a set of variables \texttt{var3} that satisfy the constraints \eqref{constraints}. This step was attempted explicitly for solutions 14 and 15 in~\cite{Andriot:2020vlg}. For solution 15, it was not successful, due the difficult lattice conditions, as discussed above. The same is true here, so we do not consider solutions with such an algebra and lattice (in particular, we also do not further pursue solutions 22-27). For solution 14, finding the set \texttt{var3} that satisfies the constraints was achieved in \cite{Andriot:2020vlg}, at the cost of lowering $r_{min}$ to $0.1$. However, as mentioned previously, factors of $2\pi$ had been forgotten in~\cite{Andriot:2020vlg}. We hence repeat the analysis here to provide a correct classicality check. The $2\pi$ factors push us in the wrong direction, meaning that one has to get to an even lower $r_{min}$. This work is carried in {\tt MSSS classicality check - sol 14.nb}, and the result is
\begin{equation}\label{sol14var3}	
\begin{aligned}
&\text{Solution 14 (\texttt{var3}):}\\[8pt]
&g_s = 0.642193\ ,\  r_1 = 4.692304\ ,\ r_2 = 9.147285\ ,\  r_3 = 0.055553\ ,\ r_4 = 6.838302\ ,\\[6pt]
&r_5 = 57.775935\ ,\ r_6 = 0.092300\ ,\ \lambda = 0.573238\ ,\\[6pt]
&N_{s1} = 16\ ,\  N_{s2} = -16\ ,\  N_{s3} = -10\ ,\ N_{1\,  5} = 43\ ,\  N_{\omega_1} = -1\ ,\  N_{\omega_2} = -1\ ,\\[6pt]
&N_1 = -0.025080\ ,\ N_2 = -0.028183\ ,\  N_3 = -1/N_2\ ,\ N_6 = -1/N_1\ ,\\[6pt]
&g^{12} = 1.206882\ ,\ g^{34} = -3.306864\ ,\ g^{56} = 1.256171\ ,\  b = -0.340830\ ,\\[6pt]
&a_{12} = -0.820253\ ,\  a_{34} = 2.087707\ ,\ a_{56} = 0.554482 \ .
   \end{aligned}
\end{equation}
One can check that all constraints \eqref{constraints} are satisfied, except for those on $r_3$ and $r_6$, which are substringy. This prevents us for now from concluding that~\eqref{sol14var3} is a classical de Sitter solution.

The numerical strategy to carry out the second step is as follows. We first find a solution in terms of \texttt{var3} that satisfies the constraints \eqref{constraints} (up to relaxing the bound $r_{min}$) over the reals rather than the integers. For instance, we would simply impose $|N_{\omega_1} | \geq 1$. Once we find such a solution, we search for a neighboring solution after having rounded some of the $N$'s to the closest integer. Doing this several times successively leads us to a final solution. We refer to {\tt MSSS classicality check - sol 14.nb} for more details. This strategy is borrowed from \cite{Andriot:2020vlg}.

\medskip

The drawback of this method is precisely that it is carried out in two steps: we first search a supergravity solution, and then hope, without any control, that it will be classical by satisfying the appropriate constraints. Combining these two steps into one is the idea of the next method.

\subsection{Direct search}\label{sec:directsearch}

In this method, we combine the two steps described above. This means that we reformulate the supergravity equations in terms of \texttt{var3} using \eqref{oldnewfields} and \eqref{quantizedqtties}, and search for solutions in terms of \texttt{var3}, requiring them to satisfy directly the classicality constraints \eqref{constraints}. This would avoid the randomness of the solution found without any check of constraints. Unfortunately, this problem turns out to be too difficult numerically. One reason is that equations formulated in terms of \texttt{var3} are much more complicated than those in terms of \texttt{var1}. Indeed, while the latter are linear or quadratic in the variables, the former have quotients that make the optimizer not find existing solutions. We illustrate this numerical difficulty in the file {\tt MSSS problem illustration.nb}. There, we perform a direct search for solution in terms of \texttt{var3}, and we tailor the constraints to allow for variables in a range for which we know that a solution exists (we use solution 14~\eqref{sol14var3}). The search fails even over the reals, i.e., without imposing integer quantization. This illustrates that the constrained numerical optimization problem solvers we tried are not robust enough to find a solution.

\medskip

The optimization problem for finding~\texttt{var3} involves solving a set of non-polynomial equations and inequalities in variables that are either real numbers or integers. The numerical framework that addresses problems of this kind is called \textit{Mixed-Integer Nonlinear Programming} (MINLP). MINLP problems are ubiquitous in scientific and engineering optimization problems, and they are notoriously difficult to address. There exists a plethora of commercial and open-source tools to address these problems. We tried numerous packages but neither led to better results than the two-step search procedure. We tried the following two direct search strategies:
\begin{itemize}
\item In a first pass, solve the problem over the reals. Then round the integer variables to the closest integer (or systematically try all combinations of integers in the vicinity of the real numbers that the optimizer found for the integer variables), fix them to this value, and run the optimizer again in the hope that this shift in some of the variables did not lift the previously close-by minimum. For the vicinity search we need to define a box within which we want to try out all possible combinations, which grows for integer variables $n_i$ with range $r_i$ as $\prod_i r_i$.
\item Apply open and closed source solvers to the problem. We can mostly treat these solvers as a black box. Depending on whether the solver is open or closed source, there is more or less information available on what the solver actually does, but it is not the focus of this paper to review the different MINLP solution strategies that these solvers employ.
\end{itemize}
For the latter option, we mostly tried different algorithms and solution techniques implemented in Mathematica. We also tried the Mathematica SHOT library~\cite{Lundell:2017aaa}.

For the first option, we tried multiple solvers that can potentially find different minima. Finding solutions over the reals is a much simpler task and a plethora of tools exist. The simplest one is just using gradient descent. The advantage is that minimizing loss functions this way is common in machine learning (ML), and there are very fast, free libraries available to perform these tasks.

One of the difficulties we encountered was that our problem is a minimization problem with non-linear constraints. There are multiple strategies to deal with these constraints: one can simply ignore them in a first pass, find a minimum of the loss function, and check whether the constraints are satisfied after the fact. Since each minimization procedure via gradient descent is very fast, this can be run for many initial guesses until a minimum that satisfies the constraints is found.

Another possibility is to build the constraints into the loss function. For example, a constraint $f(v)\geq 1$ could be incorporated in the loss as
\begin{align}
\mathcal{L}\to\lambda_1\mathcal{L}+\lambda_2\max(0,1-f(v))\,,
\end{align}
where $\lambda_1$ and $\lambda_2$ are hyperparameters that weigh the two pieces of the loss. The term we added is called a Hinge loss and is a popular choice in ML classification tasks. Note that if $f(v)\geq1$, the loss is zero, and otherwise it grows linearly with $f(v)$. We tried both approaches, but found that the solver often cannot find a minimum in which all constraints are satisfied. Moreover, for the few instances where a solution was found, we saw that rounding to the nearest integer lifts the minimum and the solution goes away.

Since it proved difficult to find a feasible starting point, i.e., an initial guess for which the constraints are solved, we also tried to tackle the problem in the opposite order: we solve the constraints without minimizing the function and then run the minimization procedure from a known good starting point. One could iterate this procedure for multiple good starting points and hope that the minimization procedure finds a nearby minimum for which the constraints are still satisfied. Another possibility is to run the minimizer with the constraints built into the loss as explained above. This approach produced the best results. We mostly used \texttt{tensorflow} for our code. We also tried the TensorFlow Constrained Optimization (TFCO) library~\cite{TFCO:2022aaa}, as well as multiple solvers that are provided in the \texttt{scipy.optimize} library, such as \texttt{basinhopping}, \texttt{BFGS}, \texttt{COBYLA}, \texttt{dual\_annealing}, \texttt{shgo}, \texttt{SLSQP}, and a simple genetic algorithm. In the past, problems like this have been tackled successfully with reinforcement learning~\cite{Halverson:2019tkf, Larfors:2020ugo, Constantin:2021for, Cole:2021nnt}, but this is very time-consuming to setup and train and we have not attempted it here.

It is interesting to ask why these methods fail. First, since we find many different minima depending on a random starting point, we know that the loss landscape is very rugged with many local extrema. The fact that the rounding approach or the box search lifts the minima is consistent with the observation that the constraints are very hard to satisfy: the constraints cut away a huge portion of the loss landscape. If these de Sitter constructions are in the swampland, they will cut away all regions that contain global minima. If global minima satisfying the constraints do exist, their basin of attraction is very small, such that the solver cannot find them most of the time. It is interesting that the authors of~\cite{Dubey:2023dvu}, who were studying flux landscapes in type IIB supergravity did not run into the problem of lifting minima when rounding. It might be insightful to study the  difference in our setups to understand which part of the geometry (solvmanifold vs Calabi--Yau) or which constraints introduce the behavior we observe but they do not.

\subsection{Constrained supergravity search}\label{sec:constrsugrasearch}

Since the equations in terms of \texttt{var3} are much more complicated than those in terms of \texttt{var1}, we also translated the constraints from \texttt{var3} to \texttt{var1} and performed the constraint optimization in terms of \texttt{var1}. However, since there are more variables in \texttt{var3} (which are related by constraints) than in \texttt{var1}, we need to find appropriate combinations to eliminate them.

In the following, we consider for simplicity the minimal value $|N_{\omega_1}|=|N_{\omega_2}|=1$. From the constraints \eqref{constraints} and the relation between \texttt{var2} and \texttt{var3} \eqref{quantizedqtties}, we obtain the following relevant combinations of variables in \texttt{var2} and conditions among them:
\begin{align}
& \frac{g_s T_{10}^1}{g_s|F_{1\, 5}| \sqrt{|f^{1}{}_{64} f^{6}{}_{14}|}} < \frac{6\times 16}{2\pi\, r_{min}^2} \ , \quad \frac{g_s T_{10}^1}{\sqrt{|f^{2}{}_{35} f^{3}{}_{25} f^{1}{}_{64} f^{6}{}_{14}|}} < \frac{6\times 16}{(2\pi)^2\, r_{min}^2}\ ,\\
& \frac{g_s |F_{3\, \omega_2}|}{g_s |F_{1\, 5}|} < \frac{1}{r_{min}^2} \ , \quad \frac{g_s |F_{3\, \omega_1}|}{ \sqrt{|f^{1}{}_{64} f^{6}{}_{14}|}} < \frac{1}{r_{min}^2} \ ,\quad \frac{g_s |T_{10}^{1,2,3}|}{g_s |F_{3\, \omega_1}|\, g_s |F_{3\, \omega_2}|} > 6\, r_{min}^2\ .\nn
\end{align}
By expressing these seven constraints in terms of \texttt{var1}, we perform a direct search for supergravity solutions including the constraints in terms of \texttt{var1}. This is done in the file {\tt MSSS solution 29.nb}. As a result, we find a new solution, $s_{55}^+ 29$, which is obtained from using the bound $r_{min}=0.01$ in the above constraints. We give it here in terms of \texttt{var1}
\begin{equation}\label{sol29var1}
\begin{aligned}
&\text{Solution 29 (\texttt{var1}):}\quad {\cal R}_4^S = 0.020309\ ,\\[8pt]
&g_s T_{10}^1 = 10\ ,\ g_s T_{10}^2 = -0.079765\ ,\ g_s T_{10}^3 = -1.064125\ ,\\[6pt]
&g_s F_{1\, 5} = -0.231074\ ,\ g_s F_{135} = -0.659250\ ,\ g_s F_{136} = -0.662773\ ,\, g_s F_{146} = 0.084135 \ ,\\[6pt]
&g_s F_{235} = -0.635765\ ,\ g_s F_{236} = -0.320255\ ,\ g_s F_{246} = -0.120817\ ,\\[6pt]
& H_{125} = -0.002972\ ,\ H_{346} = -0.181872\ ,\\[6pt]
&f^1{}_{45} = 0.829116\ ,\ f^1{}_{46} = -0.837373\ ,\ f^2{}_{35} = -0.256521\ ,\ f^2{}_{45} = -0.066547\ ,\\[6pt]
&f^2{}_{46} = -0.807542\ ,\ f^3{}_{15} = -0.013195\ ,\ f^3{}_{25} = 0.013682\ ,\ f^6{}_{14} = -0.553790\ .
\end{aligned}
\end{equation}
The $r_{min}$ value that had to be used is fairly small, so we anticipate that we did not gain much on the classicality when proceeding with this method. The classicality check of this solution 29 was performed in the file {\tt MSSS classicality check - sol 29.nb} and leads to the following values for \texttt{var3}
\begin{equation}\label{sol29var3}
\begin{aligned}
&\text{Solution 29 (var3):}\\[8pt]
&g_s = 0.532758\ ,\  r_1 = 4.704542\ ,\ r_2 = 112.925701\ ,\  r_3 = 0.067605\ ,\ r_4 = 14.968801\ ,\\[6pt]
&r_5 = 172.058417\ ,\ r_6 = 0.077310\ ,\ \lambda = 1.622330\ ,\\[6pt]
&N_{s1} = 16\ ,\  N_{s2} = -67\ ,\  N_{s3} = -68\ ,\ N_{1\,  5} = -46\ ,\  N_{\omega_1} = 1\ ,\  N_{\omega_2} = -18\ ,\\[6pt]
&N_1 = 0.020207\ ,\ N_2 = -0.002592\ ,\  N_3 = -1/N_2\ ,\ N_6 = -1/N_1\ ,\\[6pt]
&g^{12} = -0.964376\ ,\ g^{34} = 3.376434\ ,\ g^{56} = -0.990140\ ,\  b = -0.217194\ ,\\[6pt]
&a_{12} = -0.772617\ ,\  a_{34} = 1.621319\ ,\ a_{56} = 0.632508 \ .
\end{aligned}
\end{equation}
The values are indeed similar to those of solution 14, so this method is not more successful in getting classical de Sitter solutions.

Since we found a new de Sitter supergravity solution, it is interesting to study its stability (see e.g.~\cite{Andriot:2021rdy}). Doing so is straightforward  with the code {\tt MSSV} of \cite{Andriot:2022bnb}, and we find one tachyon in the spectrum of solution 29, with $\eta_V=-4.36757$. Such results are very standard~\cite{Andriot:2022bnb}.

\section{Scaling and parametric control}\label{sec:scaling}

In this section, we present a scaling freedom in our 10d solution ansatz. We discuss attempts and difficulties in getting it in a corresponding 4d theory. We finally comment on its implications regarding a parametric control on classicality and scale separation. We also compare this to related results in the literature.

\subsection{Scaling in 10d}\label{sec:scaling10d}

The scaling and its effect is best seen when looking at the relation between \texttt{var2} and \texttt{var3} as given in \eqref{quantizedqtties}. We repeat these relations here and add a tilde to the quantities scaled with $\lambda$, as explained around \eqref{lambdascaling}.
\begin{align}
& g_s \tilde{F}_{1\, 5}= \frac{g_s F_{1\, 5}}{\lambda}= \frac{g_s\, N_{1\, 5}}{r_5} \ ,\ g_s \tilde{F}_{3\, \omega_1}= \frac{g_s F_{3\, \omega_1}}{\lambda}= \frac{g_s\, N_{\omega_1}}{r_1 r_4 r_6}  \ ,\ g_s \tilde{F}_{3\, \omega_2} = \frac{g_s F_{3\, \omega_2}}{\lambda} = \frac{g_s\, N_{\omega_2}}{r_2 r_3 r_5} \ ,  \nn\\
& g_s \tilde{T}_{10}^1 = \frac{g_s T_{10}^1}{\lambda^2} = \frac{6\, g_s\, N_{s1}}{r_3 r_4 r_5 r_6}\ ,\  g_s \tilde{T}_{10}^2 = \frac{g_s T_{10}^2}{\lambda^2} = \frac{6\, g_s\, N_{s2}}{r_1 r_2 r_5 r_6}\ , \  g_s \tilde{T}_{10}^3 = \frac{g_s T_{10}^3}{\lambda^2} = \frac{6\, g_s\, N_{s3}}{r_1 r_2 r_3 r_4}\ , \label{tildequant}\\
& \tilde{f}^2{}_{35} = \frac{f^2{}_{35}}{\lambda} = \frac{2\pi\, r_2\, N_{2}}{r_3 r_5} \ ,\ \tilde{f}^3{}_{25} = \frac{f^3{}_{25}}{\lambda} = \frac{2\pi\, r_3\, N_{3}}{r_2 r_5} \ ,\nn\\
& \tilde{f}^1{}_{64} = \frac{f^1{}_{64}}{\lambda} = \frac{2\pi\, r_1\, N_{1}}{r_4 r_6} \ , \ \tilde{f}^6{}_{14} = \frac{f^6{}_{14}}{\lambda} = \frac{2\pi\, r_6\, N_{6}}{r_1 r_4} \ .\nn
\end{align}
Recall that these rescaled quantities are also a 10d solution. Next, we consider the following scaling by a (possibly discrete) parameter $\gamma \geq 1$
\begin{align}
& r_{4,5} \rightarrow \gamma\, r_{4,5} \ ,\ \ r_{1,2} \rightarrow \gamma^{x_{1,2}}\, r_{1,2} \ ,\ \ N_{s2,3} \rightarrow \gamma^{x_1+x_2-1}\, N_{s2,3} \ , \ \ N_{\omega_{1,2}} \rightarrow \gamma^{x_{1,2}}\,  N_{\omega_{1,2}} \ , \label{scaling}\\
& N_1 \rightarrow \gamma^{-x_{1}}\, N_1 \ ,\ N_6 \rightarrow \gamma^{x_{1}}\, N_6 \ ,\ N_2 \rightarrow \gamma^{-x_{2}}\, N_2 \ ,\ N_3 \rightarrow \gamma^{x_{2}}\, N_3 \ ,\nn
\end{align}
with $x_{1,2}\geq 0$, $x_1+x_2\geq 1$, and $g_s, r_3, r_6, N_{s1}, N_{1\, 5}$ invariant.
This scaling leads to a scaling of the quantities in~\eqref{tildequant},
\begin{equation}
\tilde{T}_{10}^I \rightarrow \frac{1}{\gamma^2}\, \tilde{T}_{10}^I \ ,\quad \tilde{F}_{\dots},\, \tilde{f}^a{}_{bc} \rightarrow \frac{1}{\gamma}\, \tilde{F}_{\dots},\, \tilde{f}^a{}_{bc} \ , \label{gammascaling}
\end{equation}
which is the same as a $\lambda$ scaling, i.e.~it actually corresponds to
\begin{equation}
\lambda \rightarrow \gamma\, \lambda  \ .\label{gammalambda}
\end{equation}
Another way to see this is that the $\gamma$ scaling \eqref{scaling} together with \eqref{gammalambda} leaves \eqref{quantizedqtties} invariant.

Since the $\gamma$ scaling \eqref{scaling} amounts to a $\lambda$ scaling, the $\gamma$-rescaled quantities still provide a solution. While some quantities scale explicitly  as in~\eqref{tildequant}, the variables $a_{ij}, b, g^{ij}$ entering \texttt{var2} or \texttt{var3} are invariant under scaling. The flux components not given in~\eqref{tildequant},  namely the $H$-flux and the non-harmonic piece of $F_3$, all contain one exterior derivative together with invariant quantities, so their components in the \pb are proportional to one of the four structure constants in~\eqref{tildequant}, and hence these flux components scale implicitly with $1/\gamma$. Making use of the relations between the different variable sets, we can check that the same holds in the \ib. The $\gamma$-scaling in \eqref{scaling} has the effect to rescale all quantities in the 10d equations uniformly by $1/\gamma^2$, which makes it obvious that {\sl the rescaled quantities still provide a solution.}

We can furthermore see that the constraints \eqref{constraints} for a classical string theory background are preserved by the $\gamma$ scaling \eqref{scaling}. First note that under the $\gamma$ scaling, radii and flux numbers increase, while the lattice conditions $N_1 N_6 = N_2 N_3 = -1$ are invariant. This is compatible with the constraints, up to the discretization of $\gamma$ necessary to preserve the integer condition of the flux numbers. A second subtlety is the following: for $x_1+x_2 > 1$, $|N_{s2}|$ and $|N_{s3}|$ increase with the scaling, so we need $N_{s2} \leq - 1$ ($N_{s3}\leq -1$ is automatic due to the source quantization and the requirement $T_{10}^3 \leq 0$), i.e., we need a larger contribution from $D_5$-branes to fit within the positive upper bound on this source number. Note that $N_{s2}\leq-1$ (or $T_{10}^2 <0$) is satisfied for solution 14 or 29. For $x_1+x_2=1$, the source numbers are unchanged and we do not need to worry about them. With this possible sign restriction and up to its discretization, any $\gamma$ scaling is {\sl compatible with constraints for a classical solution.}

One simple example for a $\gamma$ scaling is
\begin{equation}
r_{1,4,5} \rightarrow \gamma\, r_{1,4,5} \ ,\ N_{\omega_{1}} \rightarrow \gamma\,  N_{\omega_{1}} \ ,\ N_1 \rightarrow \gamma^{-1}\, N_1 \ ,\ N_6 \rightarrow \gamma\, N_6 \ ,\label{scalingsimple}
\end{equation}
where the constraint that $N_{\omega_{1}}\in\mathbbm{Z}$ requires discrete values for $\gamma$. Another, more balanced, example, which will be useful in the following, is given by
\begin{align}
& r_{4,5} \rightarrow \gamma\, r_{4,5} \ ,\ \ r_{1,2} \rightarrow \gamma^{\frac{1}{2}}\, r_{1,2} \ , \ \ N_{\omega_{1,2}} \rightarrow \gamma^{\frac{1}{2}}\,  N_{\omega_{1,2}} \ , \label{scalingbalanced}\\
& N_1 \rightarrow \gamma^{-\frac{1}{2}}\, N_1 \ ,\ N_6 \rightarrow \gamma^{\frac{1}{2}}\, N_6 \ ,\ N_2 \rightarrow \gamma^{-\frac{1}{2}}\, N_2 \ ,\ N_3 \rightarrow \gamma^{\frac{1}{2}}\, N_3 \ .\nn
\end{align}
Before commenting on the physical effect of this scaling, let us discuss its realization in 4d.

\subsection{Scaling in 4d}\label{sec:4dscaling}

In this subsection, we take first steps towards realizing the $\gamma$ scaling \eqref{scaling} via a combined action on scalar fields in a 4d effective theory and the 10d background quantities that enter the 4d scalar potential. A full-fledged analysis would require working out the map between the 10d variables \texttt{var3} and the 4d scalar fields, which we leave for future investigation. Our findings are presented in the file {\tt MSSV scaling attempts.nb}.

We consider a 4d theory obtained from a consistent truncation of 10d type II supergravity discussed in~\cite{Andriot:2022bnb}. The equations of motion in 4d are equivalent to our 10d equations. Using the code {\tt MSSV.nb}, it is straightforward to get this theory, in particular its scalar potential for any of our 10d solutions. For the sake of being explicit, we use solution 29 in the following, even though only the solution ansatz and not the explicit numerical values matter in the derivation. The theory has 22 scalar fields (see \cite[(2.4)]{Andriot:2022bnb}), including 6 diagonal metric fluctuations $g_{ii}$, 3 off-diagonal ones $g_{12}, g_{34}, g_{56}$, the dilaton, and flux axions. For simplicity in most of our analysis, the 12 flux axions will be set to 0, which is their background value. Also, we never really consider the dilaton since $g_s$ is left invariant under the $\gamma$ scaling.

We first consider the solution ansatz in terms of \texttt{var1}, which fixes the 4d scalar potential. We then express the scalar potential in terms of \texttt{var2} by replacing the background quantities. In this (off-shell) potential, a first test is then to scale the \texttt{var2} quantities as in \eqref{gammascaling}, which leads to a scaling of the entire scalar potential with $\frac{1}{\gamma^2}$, as expected.

We now want to reproduce the ``microscopic'' version of the above, namely the $\gamma$ scaling presented in \eqref{scaling}, by a combined action on 4d fields and background quantities. More precisely, the radii should be traded for 4d scalar fields, while the flux, source and structure constant numbers should correspond to the background quantities entering the scalar potential. Looking at the way structure constants appear in the quantity $\alpha$ or in the harmonic forms $\omega_i$, a scaling with $x_1=x_2$ seems easier to accommodate. Picking $x_i=\frac12$ also allows to scale the source volumes in the same way, while the source numbers do not scale. So we decide to try to implement the ``balanced'' scaling \eqref{scalingbalanced}. Regarding radii and 4d scalar fields, a natural guess is to consider the diagonal metric fields $g_{ii}$ to go as radii square. Since the scaling of radii is now taken care of by that of 4d fields, we propose that the background flux components and structure constants inherit their scaling properties from $N_{\omega_i}$ and $N_j$, respectively. In short, we consider the combined scaling of 4d fields and background quantities
\begin{align}
& g_{11} \rightarrow \gamma\, g_{11},\, g_{22} \rightarrow \gamma\, g_{22},\, g_{44} \rightarrow \gamma^2\,  g_{44},\, g_{55} \rightarrow \gamma^2\,  g_{55}\ ,\nn\\
& F_{3\omega_1} \rightarrow \gamma^\frac12  F_{3\omega_1},\, F_{3\omega_2} \rightarrow \gamma^\frac12  F_{3\omega_2}\ , \label{scaling4d}\\
& f^1{}_{46} \rightarrow \gamma^{-\frac12} f^1{}_{46},\, f^6{}_{14} \rightarrow \gamma^\frac12 f^6{}_{14},\, f^2{}_{35} \rightarrow \gamma^{-\frac12}   f^2{}_{35},\, f^3{}_{25} \rightarrow \gamma^\frac12 f^3{}_{25}\ .\nn
\end{align}
Ignoring the other \texttt{var2} solution parameters $b,a_{12},a_{34},,a_{56}, g^{12}, g^{34},g^{56}$ for now, as well as the other scalar fields (in particular the off-diagonal metric components that we set to 0), we obtain that the potential scales uniformly as $\frac{1}{\gamma^5}$. As we will explain, this is the expected scaling, so this uniform scaling confirms the above guesses. The difference between the $\frac{1}{\gamma^2}$ scaling of the 10d action and the $\frac{1}{\gamma^5}$ scaling of the 4d scalar potential, is due to the change between string frame used in 10d and the Einstein frame used in 4d. Going from the former to the latter, ${\cal R}_4^S$ becomes ${\cal R}_4$ and the difference is a factor $e^{2\phi} / {\rm vol}$, which also appears as an overall factor in the scalar potential. The scaling of the radii (or the $g_{ii}$) means that the 6d volume ${\rm vol}$ scales with $\gamma^3$, which explains the scaling of the scalar potential and ${\cal R}_4$ with $\frac{1}{\gamma^5}$.

Let us now consider the other flux components. For instance, the $H$-flux is given in both the \ib and \pb by
\begin{equation}
H = b \left( f^3{}_{25} e^{125} - f^1{}_{46} e^{346} \right) =  b \left( f^3{}_{25} {e^{125}}' - f^1{}_{46} {e^{34}}'\w ({e^6}' - g^{56}{e^5}' ) \right) \ .
\end{equation}
The quantity that enters the 10d as well as 4d equations is the square of each component. The 10d $\gamma$ scaling and the 4d version considered in \eqref{scaling4d} give different results: first, the structure constants entering the flux components scale differently. Second, the inverse metric entering the squares scale in 4d. More precisely, considering for instance the first $H$-flux component, one gets the following contribution in the scalar potential
\begin{equation}
|H_{125}|^2  = b^2 (f^3{}_{25})^2 g^{11} g^{22} g^{55} \ .
\end{equation}
The 4d scaling \eqref{scaling4d} makes this scale as $1/\gamma^3$. With the overall volume factor, this leads to a $1/\gamma^6$ scaling for this potential term. Therefore, to restore the right scaling, one reads off that $b \rightarrow \gamma^{\frac{1}{2}} \, b$. With similar arguments, we eventually find that we need
\begin{equation}
b \rightarrow \gamma^{\frac{1}{2}} \, b \ ,\ a_{12}\rightarrow \gamma \, a_{12}\ ,\ a_{34}\rightarrow \gamma^{-1} \, a_{34}\ ,\ a_{56} \rightarrow \gamma^{-1} \,  a_{56} \ .
\end{equation}
Combining this with \eqref{scaling4d}, and ignoring the other solution parameters $g^{12}, g^{34},g^{56}$ as well as the rest of 4d fields, we obtain the correct overall scaling of the scalar potential.

\medskip

We end with a discussion of the contribution of the 10d solution parameters $g^{12}, g^{34},g^{56}$. Since those are related to the off-diagonal metric in the \pb, we also incorporate the 4d off-diagonal metric component fields. However, we fail to find an appropriate scaling of $g^{12}, g^{34},g^{56}$ that would provide the desired scaling of the scalar potential. We tried both in the \ib (up to a change of variables towards \texttt{var2}, as explained above), where the 4d off-diagonal metric components are extremized at 0, and in the \pb (see Appendix~\ref{ap:changebasis}), where their extremal value is related to the 10d solution parameters $g^{12}, g^{34},g^{56}$. Off-shell, the volume along the 56-direction is given in terms of the 4d fields by $\sqrt{g_{55} g_{66}- g_{56}^2}$, and such combinations appear frequently in the potential. It is then natural to require the scaling of the field $g_{56} \rightarrow \gamma\, g_{56}$, to ensure a homogeneous scaling, and similarly for the other 4d off-diagonal metric fields. This, however, does not allow to get the desired scaling of the potential. Our (incorrect) guess was based on intuition gained from a diagonal metric. We would hence need the explicit map between the 10d variables \texttt{var3} (in particular the radii) and the 4d fields. Let us finally add that we made further attempts, not documented, such as evaluating the rescaled potentials on-shell, to see if we could at least find some scalings that work at an extremum, but such attempts were unsuccessful as well.

\subsection{Physical effect: parametric control on classicality and scale separation?}\label{sec:parcontrol}

The $\gamma$ scaling presented in \eqref{scaling} maps a supergravity solution to another supergravity solution, and it is compatible with constraints for a classical string background (up to a discretization of the parameter $\gamma$, and a possible sign condition mentioned above). It is then natural to wonder whether the $\gamma$ scaling can play the role of a parametric scaling that improves the classicality and/or the scale separation of a solution parametrically as $\gamma \rightarrow \infty$.

The prime observation is that a generic $\gamma$ scaling increases the size of four of the radii, while leaving the other two radii $r_3$ and $r_6$ (as well as the coupling $g_s$) invariant. This improves classicality, since in particular the 6d volume grows parametrically with $\gamma$. Hence, as long as the values of the unscaled quantities are appropriate for a classical solution, we would have parametric control on classicality. In the solutions considered above, we obtained
\begin{align}
\text{Solution 14:} \quad & g_s \approx 0.64\ ,\  r_3 \approx 0.05\ ,\ r_6 \approx 0.09\ ,\\
\text{Solution 29:} \quad & g_s \approx 0.53\ ,\  r_3 \approx 0.07\ ,\ r_6 \approx 0.08\ .
\end{align}
Those values could be considered valid for a classical solution regarding $g_s$, but problematic for $r_3$ and $r_6$, which are substringy. It is actually non-trivial to assess whether solutions with such radii are really ``non-classical''. They would spoil classicality if they can carry stringy degrees of freedom. Since these radii are actually characteristic length scales in the non-closed one-forms $e^3$ and $e^6$, they do not correspond to actual cycles in the 6d manifold. It is thus unclear to us whether they would carry stringy winding modes. In addition, various corrections would need to be evaluated. This question deserves deeper investigation, and we come back to it in the Outlook.

The fact we do not get a parametric control on some radii and on $g_s$ may also be perceived by some as a symptom of a general obstruction to get (classical) de Sitter solutions from string theory. In our compactification ansatz, we can illustrate the obstruction to scaling as follows. Using the relations \eqref{quantizedqtties} and then the constraints \eqref{constraints}, we get
\begin{align}
& \frac{g_s T_{10}^1}{g_s|F_{1\, 5}| \sqrt{|f^{1}{}_{64} f^{6}{}_{14}|}} = \frac{6 N_{s1}}{2\pi |N_{1\, 5}| \sqrt{|N_1 N_6|}}\, \frac{1}{r_3 r_6}  \ , \nn\\
& \frac{g_s T_{10}^1}{\sqrt{|f^{2}{}_{35} f^{3}{}_{25} f^{1}{}_{64} f^{6}{}_{14}|}} = \frac{6 N_{s1}}{(2\pi)^2  \sqrt{|N_2 N_3 N_1 N_6|}}\ \frac{g_s}{r_3 r_6} \nn\\
\Rightarrow\quad  & r_3 r_6 \leq \frac{6\times 16}{2\pi}\ \frac{g_s|F_{1\, 5}| \sqrt{|f^{1}{}_{64} f^{6}{}_{14}|}}{g_s T_{10}^1} \ ,\qquad g_s \geq \frac{(2\pi\, r_{min})^2}{6\times 16}\ \frac{g_s T_{10}^1}{\sqrt{|f^{2}{}_{35} f^{3}{}_{25} f^{1}{}_{64} f^{6}{}_{14}|}} \ ,
\end{align}
which shows that given a supergravity solution, there is an upper bound on $r_3 r_6$ and a lower bound on $g_s$. Hence, they  cannot be scaled parametrically in the classical regime, which agrees with the fact that they are invariant under the $\gamma$ scaling. It would be interesting to promote this to a no-go theorem or at least extend this statement to other compactification settings, but we do not see a direct generalization of the above. Note that for our solutions, we obtain
\begin{align}
\text{Solution 14:} \quad & r_3 r_6 \leq 0.220485 \ , \ g_s \geq 125.245532 \times r_{min}^2 \ ,\\
\text{Solution 29:} \quad & r_3 r_6 \leq 0.240422 \ , \ g_s \geq 101.934270 \times r_{min}^2 \ ,
\end{align}
which explains the small $r_3 r_6$ and $r_{min}$ that we obtained.

On the other hand, it has been argued in \cite[Sec. 5]{Andriot:2019wrs} (see also \cite{Junghans:2018gdb}) that hierarchies in the internal space, e.g.~among the radii, are necessary to get a classical de Sitter solution. The above parametric scaling of some radii but not of others provides a realization of such a hierarchy. If $r_3, r_6$ had been bigger, we could have claimed having a classical solution. Whether substringy values are mandatory is unclear to us at this stage.

\medskip

Having parametric control over classicality in a de Sitter solution (provided $r_3, r_6, g_s$ get appropriate values) seems in contradiction with the analysis of \cite{Junghans:2018gdb}. This work focuses on type IIA compactifications with $O_6$-planes (with possible generalizations). At first, a combination of the volume and dilaton fields is assumed as the ``classical'' direction along which one has parametric control. It is then shown that in this context (with a 4d scalar potential) the existence of such a direction is incompatible with a de Sitter solution. This does not apply to the situation above, where we only have specific radii that get scaled. In a would-be scalar potential, one would need to consider the dependence of each term on the relevant radii to get the correct scaling; the dependence on the volume alone is not enough, even if the volume also gets parametrically large. Reference \cite{Junghans:2018gdb} then describes a more general analysis in which the ``classical'' direction corresponds to any field becoming large. This situation is closer to our scaling as we will explain. Let us then briefly recall the corresponding argument adapted to our setup.

A scalar potential from a classical compactification typically receives the following contributions (we drop numerical factors and indicate only the origins and signs of the contribution)
\begin{equation}
V \sim - \sum_I T_{10}^I + |H|^2 + \sum_q g_s^2 |F_q|^2 - {\cal R}_6 \ , \label{Vapprox}
\end{equation}
where we recall that a de Sitter solution requires ${\cal R}_6 < 0$ and $\sum_I T_{10}^I > 0$ (see e.g.~\cite{Andriot:2022xjh}). The potential can then typically be written in terms of a field $r>0$ as
\begin{equation}
V = - \frac{A_{O_p}}{r^2} + \sum_i \frac{A_i(r)}{r^2} \ , \label{Vr}
\end{equation}
where the (dominant) constant orientifold contribution $A_{O_p}$ and the other contributions $A_i(r)$ satisfy $A_{O_p}, A_i(r) \geq 0$. The signs and the dependence on $r$, could be argued for more generally, as e.g.~in \cite{Junghans:2018gdb}. We content ourselves here with verifying them in our setup. The signs can already be read off from \eqref{Vapprox}, together with the fact we have $T_{10}^1 >0$, $T_{10}^{2,3}<0$. Using the potential \eqref{Vr} and the equation $\del_r V = 0$, we get an extremum at
\begin{equation}
2 r V = \sum_i \del_r A_i \ ,
\end{equation}
from which one concludes that a de Sitter solution requires $\del_r A_i >0$ for at least some $i$ at the extremum. Under the hypothesis (which holds in our case) that $A_i(r) \sim \frac{B_i}{r^{y_i}} \geq 0$, in the limit $r \rightarrow \infty$, we conclude that $\del_r A_i = - y_i\, \frac{B_i}{r^{y_i+1}}$ has the sign of $-y_i$. Typically, one has $y_i \geq 0$; if this is true for all $y_i$, then one cannot have a de Sitter solution at large $r$. Let us examine how our setting circumvents this argument of \cite{Junghans:2018gdb}.

We consider the simple scaling relations~\eqref{scalingsimple}, and introduce a single radius $r$ with which $r_1,r_4,r_5$ scale simultaneously. For simplicity, we also only consider the harmonic components of the fluxes. This results in
\begin{align}
& g_s^2 F_{1\, 5}^2 \sim \frac{1}{r^2} \ ,\ g_s^2 F_{3 \omega_1}^2 \sim \frac{N_{\omega_1}^2}{r^4} \ ,\ g_s^2 F_{3 \omega_2}^2 \sim \frac{1}{r^2} \ ,\ g_s T_{10}^1 \sim \frac{1}{r^2} \ ,\ g_s T_{10}^{2,3} \sim -\frac{1}{r^2} \ , \label{Circumr1}\\
& f^2{}_{35} \sim \pm \frac{1}{r} \ ,\ f^3{}_{25} \sim \mp \frac{1}{r} \ ,\ f^1{}_{64} \sim \frac{r N_1}{r} \ ,\ f^6{}_{14} \sim \frac{N_6}{r^2} \ .\label{Circumr2}
\end{align}
The contributions from \eqref{Circumr1} satisfy all assumptions made above, with $y_i=0,2$, thus not helping to get a de Sitter extremum. The curvature contribution in the \pb can be read off from the formula
\begin{equation}
- 2 {\cal R}_6 = (f^2{}_{35} + f^3{}_{25})^2 + (f^1{}_{64} + f^6{}_{14} )^2 \ .
\end{equation}
We obtain the correct signs, and the question comes down to considering the $r$ dependence. From $(f^2{}_{35} + f^3{}_{25})^2$, we also get $y_i=0$, but
\begin{equation}
\label{eq:CurvatureScaling}
(f^1{}_{64} + f^6{}_{14} )^2 \sim \frac{1}{r^2} \left( \frac{N_6}{r} + r N_1 \right)^2 \sim \frac{1}{r^2} \times (  r N_1 )^2 \ ,
\end{equation}
which gives $y_i=-2$ at large $r$. This term allows to circumvent the argument of \cite{Junghans:2018gdb}.

This possibility is actually mentioned as a loophole in \cite{Junghans:2018gdb}, in the case where $A_i \sim \epsilon \, r^{-y}$ with $y<0$ and $\epsilon \rightarrow 0$. This is precisely realized in our setup with $y=-2$, $\epsilon = N_1^2 \sim \gamma^{-2}$ for $r^2 \sim \gamma^2$. It is suggested in \cite{Junghans:2018gdb} that $\epsilon$ could be due to a warp factor, but that structure constants tend to have $N_i \sim {\cal O}(1)$. While the latter is true for nilmanifolds with integer $N_i$ (a condition easily obtained with the same method as in section \ref{sec:quant}), we see that solvmanifolds offer different possibilities for the structure constants due to the different lattice quantization conditions. This difference for the structure constants, or equivalently of the solvmanifold fibration and compactness conditions, allows for the $\gamma$ scaling.

It might seem puzzling there seemingly is a continuous scaling degree of freedom in the geometry which acts as a breathing mode and could be an unfixed modulus.\footnote{We thank Daniel Junghans for a related discussion. More discussions on the 6d geometry can be found around \eqref{lattices}; we also refer to \cite{Andriot:2010ju, Grana:2013ila, Andriot:2015sia, Acharya:2019mcu, Acharya:2020hsc, Andriot:2020vlg} for other works where this geometry appeared.} Let us first recall that all 10d equations have been solved to find these solutions, so it is unlikely that some geometric modulus capturing this freedom has been missed. Focusing on the structure constants, it is easy to see that this scaling freedom is nothing but the tuning of the ratio between radii, $r_1/r_6$ or $r_2/r_3$. Indeed, the $\gamma$ scaling changes the $N_i$ by the same amount as those ratios. More precisely, the change of these ratios can be implemented by scaling one of the radii while having the second one fixed. This is the freedom explicitly used in \cite{Grana:2013ila} to get $|N_1|=|N_6|$, a situation we can also reach here with the $\gamma$ scaling. In our case, we would obtain $|N_1|=|N_6|=1$.\footnote{Let us emphasize once again that compactness, for solvmanifolds, does not enforce $|N_1|=|N_6| \in \mathbb{Z}$; it only quantizes the product $|N_1 N_6|$.} We also recall that the metric has determinant one in either basis, which is why the radii enter in the structure constants and the scalings of $r_1$, or $r_1/r_6$, compensates that of $N_1$ and $N_6$. Once we solve the 10d equations, the values of the $N_i$ and $r_i$ are actually fixed and no such scaling freedom is left. One can then perform the $\gamma$ scaling, but this maps it to a different solution (for instance with different ${\cal R}_4$, flux number, etc.).

In addition, note that the scaling freedom acting on actual solutions is not continuous but discrete to ensure proper flux quantization. Even in the simplest scaling example \eqref{scalingsimple}, we see all these essential features together: scaling of the $N_i$ in the structure constants, together with that of the radii and a scaling of the flux number which imposes discretization of the parameter. The scaling may be viewed as analogous to what happens in the DGKT solution \cite{DeWolfe:2005uu}: the parameter there is a (discrete) flux integer, a formulation we could also use here. Also there, ${\cal R}_4$ or the scalar potential scale uniformly in the parameter, as it does here, and the radii also get rescaled.

\medskip

Let us also briefly compare our result to the obstruction on asymptotic de Sitter solutions presented in \cite{Grimm:2019ixq} (the remarks to come also apply to \cite{Cicoli:2021fsd}), which considers compactifications on Ricci-flat manifolds. In contrast, the curvature term~\eqref{eq:CurvatureScaling} plays a crucial role in allowing a de Sitter solution with parametric scaling in our case. While dualities could potentially map the obstruction of \cite{Grimm:2019ixq} to settings with curvature, this does not include our setup. Solvmanifolds of the kind considered here, with a pair $f^1{}_{64}, f^6{}_{14}$, would map under the naive T-duality rules of \cite{Shelton:2005cf}, to a space without structure constants but with non-trivial $H$-flux ($H_{164}$) together with non-geometric $Q$-flux, $Q_1{}^{64}$ (T-duality along direction 1) \cite{Hassler:2014sba, Andriot:2015sia}. Since the compactifications of \cite{Grimm:2019ixq} do not capture non-geometric fluxes, it seems that our setting circumvents the obstruction described there.\footnote{The $\gamma$ scaling allows the structure constants $f^1{}_{64}, f^6{}_{14}$ (or more precisely $N_1,N_6$) to take non-integer values, and this would result in a non-quantized $H_{164}$. This can be explained by the necessary presence of the non-geometric $Q_1{}^{64}$, which makes the identification of a standard differentiable manifold challenging; as a result, it is not clear that the $H$-flux is on a cycle with an associated quantization condition. Interestingly, this can be contrasted with the nilmanifold case mentioned above, where the (non-paired) structure constant has to be given by an integer, and the T-dual (geometric) setting would have a quantized $H$-flux.} Finally, the compactifications considered here make it difficult to formulate the corresponding 4d theory as a standard ${\cal N}=1$ 4d supergravity (see \cite[Sec. 6]{Andriot:2022bnb}). Therefore, arguments using the ${\cal N}=1$ superpotential and K\"ahler potential may not apply.

\medskip

The analogy with the DGKT solution brings us to one last aspect of the $\gamma$ scaling: the question of scale separation. In our 6d geometry, the only free circles are along directions $4,5$ (the others are fibered). Those lead to Kaluza--Klein (KK) masses given in the 10d string frame by integer multiples of $1/r_{4,5}$. This leads to a scaling $m_{\text{KK},\, S}^2 \sim 1/\gamma^2$. To investigate scale separation, this should be compared to ${\cal R}_4^S$ in the 10d string frame, which scales in the same way, ${\cal R}_4^S \sim 1/\gamma^2$. In the 4d Einstein frame, ${\cal R}_4$ goes as ${\cal R}_4 \sim 1/\gamma^5$ (for $x_1+x_2=1$). However, the 4d Einstein frame KK masses $m_\text{KK}^2$ are modified by the same factor, which arises from the inverse 4d metric. Thus, at first sight there is no scale separation, i.e., in terms of the $\gamma$ scaling we have
\begin{equation}
\frac{m_\text{KK}^2 }{ {\cal R}_4 } \sim 1 \ ,
\end{equation}
saturating the bound proposed in \cite{Lust:2019zwm}. This is consistent with the belief that for a de Sitter solution, at best a local or numerical scale separation can occur, not a parametric one, due to the plethora of constraints and bounds that have to be obeyed \cite{Andriot:2020vlg}. As an illustration, it is then worth to compute the numerical values obtained for these quantities, and we do so in the 10d string frame for solutions 14 and 29. Recall that the value of the 4d Ricci scalar is given in units of $2\pi l_s$. The lowest KK mass of the tower is given by $1/r_{5}$ (since $r_4$ is about 10 times smaller in both solutions, giving a higher mass), and the radii are given in units of~$l_s$. Converting them to units of $2\pi l_s$ and applying the appropriate $\lambda$ rescaling to the Ricci scalar, we get
\begin{align}
\hspace{-0.1in} \text{Solution 14:} \quad & \tilde{{\cal R}}_4^S = \frac{{\cal R}_4^S}{\lambda^2} = 0.068953 \ (2\pi l_s)^{-2} \ ,\ m_{\text{KK},\, S}^2 = \frac{(2\pi)^2}{r_5^2} = 0.011827 \ (2\pi l_s)^{-2}  \ ,\\
\hspace{-0.1in} \text{Solution 29:} \quad & \tilde{{\cal R}}_4^S = \frac{{\cal R}_4^S}{\lambda^2} = 0.007716 \ (2\pi l_s)^{-2} \ ,\ m_{\text{KK},\, S}^2 = \frac{(2\pi)^2}{r_5^2} = 0.001334 \  (2\pi l_s)^{-2}  \ ,
\end{align}
which does not exhibit any obvious numerical separation, even when taking into account a factor of 4 to reach a cosmological constant. Note that the tower with $r_4$ would, on the other hand, have an interesting numerical mass gap. There remains one possible loophole to this apparent absence of scale separation: there could be other contributions to the KK spectrum, arising from the rest of the geometry, as well as from the fluxes and sources. We have not computed these contributions here. Note that the same also applies to DGKT (see e.g.~\cite[App. D]{Caviezel:2008ik}). However, since all supergravity quantities scale with $1/\gamma^2$ (in particular the structure constants which characterize the rest of the geometry) in our case, it could be that all $m_{\text{KK},\, S}^2$ scale in the same way and thus do not lead to parametric scale separation. Numerical gaps would still have to be computed.

\section{Outlook}\label{sec:outlook}

In this paper, we focused on a 10d type IIB supergravity ansatz for a de Sitter solution. We developed and discussed numerical methods to find such solutions subject to constraints that ensure that such solutions are classical string backgrounds. Furthermore, we identified an interesting scaling freedom in our ansatz, which maps one solution to another while preserving the classicality constraints. Moreover, this scaling can make 4 radii parametrically large. If the other 2 radii, $r_3, r_6$, and the string coupling constant, $g_s$, which are left invariant by the scaling, turned out to have appropriate values, we could obtain a classical solution with parametric control. However, such a solution is unlikely to be scale separated, at least parametrically. The numerical search for solutions has given two solutions $s_{55}^+ 14$ and $s_{55}^+ 29$, given in \eqref{sol14var1}-\eqref{sol14var3}, and \eqref{sol29var1}-\eqref{sol29var3}, respectively. In both solutions we get $g_s<1$, but $r_3$ and $r_6$ are of substringy lengths. It is unclear to us whether the latter is a general feature in this ansatz (or even beyond it), or whether getting better solutions is possible but numerically challenging; at least we see hints of such numerical difficulties. We refer to the Introduction for a more detailed summary of our results.

Finding isolated dimension zero objects in a high-dimensional parameter space is often numerically quite challenging. This means that numerical searches are much more difficult if asymptotic de Sitter solutions or family of solutions related by a (discrete) parameter do not exist. Besides this problem, the glassy nature of the string landscape with a huge number of extrema poses challenges to numerical minimization algorithms. Finding the correct ones among exponentially many will only be possible if the basin of attraction of the former is large enough. As we have seen in section \ref{sec:directsearch}, reformulating the problem in a different basis means that solutions that could be found previously are not found in the new basis, hinting at an ill-conditioned numerical problem. In this context one could investigate several questions: first, one could try to identify subsets of conditions that can be easily satisfied together, but adding another breaks this property. In a systematic study, one would find the power set of all constraints and run minimization attempts on all of them. This is expected to be feasible since the minimization procedure only takes a few seconds or minutes in the worst case and the problem can be parallelized. This would give insight into which constraints seemingly contradict each other and could lead to the formulation of no-go theorems. On the flip side, it might inform a change of basis or an order in which the constraints should be imposed to make the problem better behaved numerically. The minimization requires solving a mixed integer non-linear programming problem, which is notoriously difficult. In the past, reinforcement learning has been applied successfully to problems of this kind~\cite{Halverson:2019tkf,Larfors:2020ugo,Constantin:2021for,Cole:2021nnt}. Moreover, coming back to the order in which constraints should be imposed, it was observed in~\cite{Halverson:2019tkf} that the reinforcement learning algorithm could identify a beneficial order of constraints in the context of type IIA supergravity on orientifolds of toroidal orbifolds with $O_6$-planes and $D_6$-brane stacks.

Having $r_3, r_6 < 1$ in string units violates our conservative classicality conditions \eqref{classicality} or constraints \eqref{constraints}. Whether this actually renders the solution non-classical is a difficult question. First, the substringy ``radii'' appear as length scales in one-forms which are not closed: in other words, they do not a priori  correspond to volumes of one-cycles. As a consequence, it is not obvious that string winding modes can arise from them. We hope to determine the cohomology and associated volumes in future work. Second, it is also unclear whether the $\alpha'$ corrections would be dominant due to these substringy lengths. Computing these corrections would actually be an important way to assess whether the solution is classical. The $\alpha'$ corrections may only involve combinations of radii, which may turn out to give small corrections despite the substringy $r_3, r_6$. We expect this to be amplified using the scaling freedom of the solutions: since the $\alpha'$ corrections would involve various powers of supergravity quantities (such as the structure constants in Riemann and Ricci tensors, flux components, etc.), which scale with $1/\gamma$, our solutions are likely to be protected from these corrections by rescaling them with $\gamma \gg1$. This would establish parametric control. We hope to come back to this important question in future work.

As a side remark, we noted already in our solution an internal hierarchy in the 6d manifold, as advocated in \cite{Andriot:2019wrs} for a classical de Sitter solution, and reinforced by the scaling. This is similar to the dark dimension scenario \cite{Montero:2022prj}, which advocates for the existence of one large extra dimension. In our solutions 14 \eqref{sol14var3} and 29 \eqref{sol29var3}, we note that among the scaled radii, some are larger than others, and we have not even tried to make such a difference more manifest. It would be interesting to see if our solution ansatz could provide a realization of this dark dimension scenario.

Beyond the question of classicality, another question that could invalidate our de Sitter solutions is the use of smeared sources. This is a common problem in supergravity solutions with intersecting sets of parallel $O_p/D_p$ sources. One should include the backreaction of these objects, which would introduce warp factors and a varying dilaton. In the case of the DGKT solution discussed previously, where an analogous scaling offers parametric control over classicality and scale separation, recent progress to describe this backreaction has been made by using an expansion in terms of the scaling parameter \cite{Junghans:2020acz, Marchesano:2020qvg} (see \cite{Andriot:2023fss} for a recent account and more references). It would be natural to analogously include the source backreaction in our de Sitter solutions by using our scaling freedom. A generic study of such a possibility has been conducted in \cite[Sec. 4]{Junghans:2023lpo}, and could be relevant to such a task. A related question is that of the size of the hole around each backreacted orientifold: if too large, this would induce important stringy corrections that would spoil the supergravity and classical regime approximations, analogously to the singular bulk problem \cite{Carta:2019rhx, Gao:2020xqh, Carta:2021lqg}. We hope to come back to this question in the future.

An option for the scaling freedom~\eqref{scaling}, which we have not explored in depth in this work, is to allow for a parametrically large number of parallel $D_5$-branes in sets $I=2,3$: this is possible for scaling exponents obeying $x_1+x_2 \neq 1$. The realization of solution 29 in terms of \texttt{var3} presented here \eqref{sol29var3} already has 83 and 68 $D_5$ branes. If these $D_5$ are localized in stacks, having a large numbers of them could violate bounds that have been proposed (in different contexts) on the rank of gauge groups (see e.g.~\cite{Kim:2019ths, Martucci:2022krl}). One could, however, argue that a large number of $D_5$ branes would cause a large backreaction, calling the supergravity approximation into question (in case it is even valid in the smeared approximation). Nevertheless, since the volumes wrapped by the $D_5$ branes also grow parametrically with $\gamma$, the situation requires a more careful study. Those matters are also related to open string degrees of freedom, and brane calibrations (see e.g.~\cite[App. B]{Andriot:2016xvq}), that have not been taken into account in this work. We hope to return to these questions in the future.

Throughout this paper, we have focused on one specific ansatz for a de Sitter solution and studied the conditions to get a classical string background. We have seen that the analysis can be involved, in part due to the intricacies of the 6d geometry and the need to change bases there. The difficulties to satisfy the lattice conditions that ensure compactness also led us to restrict our focus to only one group manifold. It is natural to ask whether such an analysis could be conducted for other examples. For all known 10d de Sitter solutions on (compact) group manifolds, the corresponding algebras are given in Table 4 of \cite{Andriot:2022yyj}. Removing those that include an algebra $\mathfrak{g}_{3.4}^{-1}$ or $\mathfrak{g}_{6.n}^{\dots}$, whose lattice conditions are too complicated to handle, we are left, beyond the example considered here, with decomposable algebras involving $\mathfrak{so}(3) = \mathfrak{su}(2)$. The latter is reminiscent of the first de Sitter solutions found \cite{Caviezel:2008tf}. To study the classicality of these solutions with seemingly manageable algebras (in particular solutions $s_{55}^+ 20,21$), we would first need an analytic change of basis to reach the basis with only the 3 structure constants we commonly know for that 3d algebra. In all these examples, we have so far only worked out the change of basis numerically. We would also need to check whether the change of basis preserves the source volumes (see e.g.~\cite[App. A]{Andriot:2022way}). Moreover, we would need the harmonic forms. With these quantities, the analysis we performed here could then be repeated. While this is a priori doable, it requires substantial amounts of work. It is also unclear whether these setups would also feature a similar scaling freedom. We hope to come back to these questions in the future.

\vspace{0.4in}

\subsection*{Acknowledgments}

We warmly thank Luc Didier for useful exchanges and many attempts to find solutions using a genetic algorithm. We also thank Daniel Junghans, Georges Obied, George Tringas and Daniel Waldram for helpful exchanges during the completion of this work. The work of FR is supported by the NSF grants PHY-2210333, PHY-2019786 (The NSF AI Institute for Artificial Intelligence and Fundamental Interactions), and startup funding from Northeastern University.

\newpage

\begin{appendix}

\section{Source contributions}\label{ap:sources}

In this appendix, we give details on the source contribution $T_{10}^I$ in \eqref{T10I} for each set $I$ of parallel sources. The definition builds on conventions of \cite[App. A]{Andriot:2016xvq} (see \cite[App.~A]{Andriot:2023fss} for a recent related discussion). This contribution can be expressed in terms of a source current $\delta^{\bot_I}_{9-p}$, which is a $(9-p)$-form. The latter is typically introduced at the level of the source action to promote the world-volume action to a 10d action, i.e., to embed the source in the 10d spacetime. For a single $O_p/D_p$, one has
\begin{equation}
\frac{S_{DBI}}{-c_p T_p g_s^{-1}} = \int_{p+1} \d \xi^{p+1} \sqrt{\iota^*[g_{10}]} = \int_{10} {\rm vol}_4 \w {\rm vol}_{||} \w \delta^{\bot}_{9-p}
\end{equation}
where we consider a constant dilaton with $e^{\phi}=g_s$ for simplicity. We have $c_p=1$ for a $D_p$ and $c_p=-2^{p-5}$ for an $O_p$, and $ T_p\, 2 \kappa_{10}^2 = (2 \pi l_s)^{7-p}$. Moreover, $\iota^*[g_{10}]$ denotes the pull-back of the 10d target-space metric to the world-volume. In the 6d (embedding) manifold, the source is wrapping a space of volume form ${\rm vol}_{||}$, while the transverse directions have a volume form ${\rm vol}_{\bot}$ such that ${\rm vol}_{6}={\rm vol}_{||} \w {\rm vol}_{\bot}$. The current form $\delta^{\bot}_{9-p}$ is sometimes understood as localizing the source in its transverse directions. For a single source, the contribution $T_{10}$ is defined as
\begin{equation}
{\rm vol}_4 \w {\rm vol}_{||} \w \delta^{\bot}_{9-p} = {\rm vol}_{10}\, \frac{T_{10}}{p+1}\times \frac{1}{-c_p T_p 2 \kappa_{10}^2} \ ,\nn
\end{equation}
or equivalently
\begin{equation}
\frac{T_{10}}{p+1}  {\rm vol}_{\bot} = -c_p\, (2 \pi l_s)^{7-p}\ \delta^{\bot}_{9-p} \ .
\end{equation}
We now introduce the smeared version, which for a single source is
\begin{equation}
T_{10}|_s = \frac{\int {\rm vol}_{10}\, T_{10}}{\int {\rm vol}_{10}} \ .
\end{equation}
We then get
\begin{equation}
\frac{T_{10}|_s}{p+1} \times \int_{10} {\rm vol}_{10} = -c_p\, (2 \pi l_s)^{7-p} \int_{10} {\rm vol}_4 \w {\rm vol}_{||} \w \delta^{\bot}_{9-p}  = -c_p\, (2 \pi l_s)^{7-p} \int_{p+1} \d \xi^{p+1} \sqrt{\iota^*[g_{10}]} \ .
\end{equation}
Restricting to space-filling sources, we consider for simplicity that the embedding into the 4d spacetime is trivial, so we can drop the 4d parts
\begin{equation}
\frac{T_{10}|_s}{p+1}\times \int_{6} {\rm vol}_{6} = -c_p\, (2 \pi l_s)^{7-p} \int_{6} {\rm vol}_{||} \w \delta^{\bot}_{9-p}  = -c_p\, (2 \pi l_s)^{7-p} \int_{p-3} \d \xi^{p-3} \sqrt{\iota^*[g_{6}]} \ . \label{T10smeared}
\end{equation}
In the simple case where the world-volume embedding is trivial, sources wrap cycles, and ${\rm vol}_{||}, {\rm vol}_{\bot}$ are corresponding harmonic forms, equation~\eqref{T10smeared} simplifies. One obtains for the measure $\d \xi^{p-3} \sqrt{\iota^*[g_{6}]} = {\rm vol}_{||}$. The 6d integral can be split in two parts with $\int \delta^{\bot}_{9-p} = 1$. One concludes in this case that $\frac{T_{10}|_s}{p+1} = -c_p\, (2 \pi l_s)^{7-p} / \int {\rm vol}_{\bot}$. From there, by summing over the sources in one set $I$, which by definition share the same ${\rm vol}_{\bot}$, it is straightforward to reproduce the smeared contribution $T_{10}^I$ in \eqref{T10I}:
\begin{equation}
\frac{T_{10}^I}{p+1} = (2^{p-5} N_{O_p}^I - N_{D_p}^I) \,  \frac{(2\pi l_s)^{7-p}}{\int {\rm vol}_{\bot_I}} \ . \label{T10Iapp}
\end{equation}

However, in many flux compactifications, the sources do not wrap cycles in the 6d manifold~$\mmm$. One typically gets that ${\rm vol}_{\bot}$ is exact in $\mmm$ and ${\rm vol}_{||}$ is not closed. This makes their integrals over subspaces not necessarily well-defined. The integral $\int_{9-p} {\rm vol}_{\bot}$  may seem to vanish since ${\rm vol}_{\bot}$ is exact, but it actually cannot be integrated over such a subspace. Nevertheless, since all forms are constructed from the vielbeins $e^a$, they are globally well-defined objects in the 6d manifold $\mmm$. This situation illustrates the role of $\delta^{\bot}_{9-p}$ and the embedding in the 6d space, or equivalently of the pull-back to the world-volume. In this (common) situation, we face the question of how to perform the same computation as before and get the smeared source contribution.

Let us briefly discuss a standard example. Consider a 6d manifold that includes a $T^3$ with $H$-flux. This solution also includes an $O_3$-plane localized on the 6d manifold. Performing a T-duality along one circle of the $T^3$, say along direction $y^3$, one obtains a new 6d manifold where the $T^3$ has become a nilmanifold $\mathcal{N}_3$, that carries no $H$-flux. Under T-duality, the $O_3$ has become an $O_4$ that wraps the T-dualized direction. This well-known example was first considered in \cite{Kachru:2002sk}, and a detailed account can be found in \cite[Sec. 3.1]{Andriot:2015sia} with related references. The nilmanifold is a simple group manifold, whose direction 3 is a circle, fibered over a $T^2$ base along 1 and 2. Schematically, we can introduce the corresponding one-forms
\begin{equation}
e^1 = r_1 \d y^1 \ , \ e^2 = r_2 \d y^2 \ , \ e^3 = r_3  ( \d y^3 - N \, y^1 \d y^2) \ ,
\end{equation}
where the fibration is encoded in the one-form $- r_3  N \, y^1 \d y^2$. The situation described here is the same as above: in the 6d manifold, the $O_4$ wraps a direction whose volume form is along $e^3$, which is not closed, i.e., not a representative of a homology class. However, one may also argue that the world-volume of the $O_4$ is a circle, along $y^3$, whose embedding in the 6d manifold is non-trivial. As a consequence, the world-volume action $\int_{1} \d \xi^{1} \sqrt{\iota^*[g_{6}]}$ appears to be independent of the fibration. This is consistent with the fact that whatever $\delta^{\bot}_{5}$ is, it has to be proportional to $\d y^1 \w \d y^2$ or $e^{12}$, so ${\rm vol}_{||} \w \delta^{\bot}_{5}$ erases the information of the fibration, i.e., $- r_3  N \, y^1 \d y^2 \w \delta^{\bot}_{5} = 0$. Therefore, we get $\int_{1} \d \xi^{1} \sqrt{\iota^*[g_{6}]} = \int_{S^1} r_3 \d y^3 = 2\pi r_3$.

Let us apply the same ideas to the examples of interest here. Consider for instance the following one-forms, discussed in section \ref{sec:quant}
\begin{align}
{e^6}' & = r_6  \left( \cos(\sqrt{|N_1 N_6|} y^4) \d y^6 - \left|\frac{N_6}{N_1}\right|^{\frac{1}{2}} \sin(\sqrt{|N_1 N_6|} y^4) \d y^1  \right) \nn\\
{e^1}' & = r_1  \left( \left|\frac{N_1}{N_6}\right|^{\frac{1}{2}} \sin(\sqrt{|N_1 N_6|} y^4) \d y^6 + \cos(\sqrt{|N_1 N_6|} y^4) \d y^1  \right) \nn\\
{e^4}' & = r_4 \d y^4 \ ,\\
{e^2}' & = r_2 \left( \cosh(\sqrt{N_2 N_3} y^5) \d y^2  + \left(\frac{N_2}{N_3}\right)^{\frac{1}{2}} \sinh(\sqrt{N_2 N_3} y^5) \d y^3  \right) \nn\\
{e^3}' & = r_3 \left( \left(\frac{N_3}{N_2}\right)^{\frac{1}{2}} \sinh(\sqrt{N_2 N_3} y^5) \d y^2 + \cosh(\sqrt{N_2 N_3} y^5) \d y^3  \right) \nn\\
{e^5}' & = r_5 \d y^5 \ .
\end{align}
These give ${e^{16}}'= r_1r_6 \, \d y^1 \w \d y^6$ and ${e^{23}}' = r_2 r_3 \, \d y^2 \w \d y^3$. A first $O_5$ is wrapped along directions 12: in the 6d space, it has ${\rm vol}_{||} = {e^{12}}'$. As above, we can interpret this as the $O_5$ wrapping two circles along $y^1, y^2$, but those circles become part of the fibers in the 6d manifold, whose base is along 45. In that case, the volume form  of the world-volume theory would be given by $r_1 r_2 \d y^1 \w \d y^2$. From the 6d perspective, we have $\delta^{\bot}_{4} \propto {e^{3456}}' \propto {e^{36}}'$. In the 6d integral, we can then obtain ${e^{16}}'\w {e^{23}}'$, allowing to drop all fibration factors ($\cos, \cosh$ etc.). This erases the fibration information, as in the above example. From both perspectives (world-volume and 6d), we then obtain
\begin{equation}
\int_{6} {\rm vol}_{||} \w \delta^{\bot}_{4}  = \int_{2} \d \xi^{2} \sqrt{\iota^*[g_{6}]} = r_1 r_2 \int_2 \d y^1 \w \d y^2 = (2\pi)^2 r_1 r_2 \ ,
\end{equation}
where $\delta^{\bot}_{4} $ also provides the correct normalization.

\medskip

We have thus argued that for a single source, even when wrapping a non-cycle in the 6d manifold, one has
\begin{equation}
\int_{6} {\rm vol}_{||} \w \delta^{\bot}_{9-p}  = \int_{p-3} \d \xi^{p-3} \sqrt{\iota^*[g_{6}]} = (2\pi)^{p-3} \prod_{a_{||}} r_{a_{||}} \ .
\end{equation}
Together with the normalization
\begin{equation}
\int_{6} {\rm vol}_{6} = (2\pi)^6 \prod_{a=1}^{6} r_a \ ,
\end{equation}
we can rewrite \eqref{T10smeared} in the presence of a single source as
\begin{equation}
\frac{T_{10}|_s}{p+1} =- c_p \frac{1}{(2\pi l_s)^2} \prod_{a_{||}}  \frac{ r_{a_{||}} }{l_s} \times \prod_{a=1}^{6} \frac{l_s}{r_a} = - c_p \frac{1}{(2\pi l_s)^2} \prod_{a_{\bot}} \frac{l_s}{r_{a_{\bot}}}  \ .
\end{equation}
Summing over all sources in the set $I$, which share the same parallel and transverse directions, one deduces the expression for $T_{10}^I$ given in \eqref{T10I} or \eqref{T10Iapp}, with the convention ``$\int {\rm vol}_{\bot_I}$''$= vol_{\bot_I} = (2 \pi)^{9-p}\ r_{a_{1\,\bot_I}} ... r_{a_{9-p\,\bot_I}}$.

\section{On the change of basis in {\tt MSSV.nb} and solution searches}\label{ap:changebasis}

We carry out the 10d search for solutions by building on {\tt MSSS.nb}~\cite{Andriot:2022way}. The code generates all 10d equations in the \ib, using that the metric components are $\delta_{ab}$, as well as the solution ansatz corresponding to \texttt{var1}. We can then express all equations in terms of \texttt{var2} and \texttt{var3} using the relations worked out in this paper.

Note that the code {\tt MSSV.nb} \cite{Andriot:2022bnb} can compute the 4d scalar potential in either basis: first, it accepts a generic metric with components $g_{ab}$ (which now enters as 4d fields). Moreover, since the placement of sources (their world-volumes) are the same in either basis, the orientifold projections are unchanged. The only changes involve the precise solution ansatz, as summarized in \eqref{changebasissummary}: beyond the metric, the non-zero flux components are slightly different in both bases. One also needs to be careful to specify the correct extremal values for the 4d fields. We illustrate this computation in the file {\tt MSSV attempts scaling.nb}: we express the off-shell scalar potential directly in the \pb, and verify that a 10d solution in that basis (obtained via relations such as \eqref{Fprime}) is indeed an extremum of the 4d potential. We then use the potential obtained directly in the \pb to study the 4d scaling there, as described in section \ref{sec:4dscaling}.

One could also search for solutions directly in the \pb by extremizing the 4d potential. The corresponding equations in {\tt MSSV.nb} were shown to be equivalent to the 10d equations of motion \cite{Andriot:2022bnb}. It is unclear to us whether the search for supergravity solutions, possibly with constraints, would be simpler or more successful in the \pb.

\end{appendix}

\newpage

\providecommand{\href}[2]{#2}\begingroup\raggedright
\endgroup

\end{document}